\documentclass[referee,a4paper,12pt,traditabstract]{jswsc}
\usepackage{amsmath}
\usepackage{graphicx}
\usepackage{txfonts}
\usepackage{subfigure}
\usepackage{epstopdf}
\usepackage[displaymath,mathlines]{lineno}
\usepackage[authoryear,round]{natbib}
\usepackage[backref]{hyperref}
\usepackage{url}
\usepackage{soul}
\usepackage{multirow}
\usepackage{comment}
\usepackage{enumitem}
\hypersetup{colorlinks=true, citecolor=cyan, urlcolor=cyan, linkcolor=blue}

\begin{document}

\title{Forecasting Solar Energetic Proton Integral Fluxes with Bi-Directional Long Short-Term Memory Neural Networks}

\titlerunning{Forecasting SEP Flux with Deep Learning Models}
\authorrunning{M. Nedal et al.}

\author{Mohamed Nedal\inst{1}\fnmsep\thanks{Corresponding author}, Kamen Kozarev\inst{1}, Nestor Arsenov\inst{1}, 
Peijin Zhang\inst{1}}

\institute{Institute of Astronomy of the Bulgarian Academy of Sciences, 1784 Sofia, Bulgaria\\
  \inst{1}\email{\href{mailto:mnedal@astro.bas.bg}{mnedal@astro.bas.bg}}\\
  \inst{1}\email{\href{mailto:kkozarev@astro.bas.bg}{kkozarev@astro.bas.bg}}\\
  \inst{1}\email{\href{mailto:narsenov@nao-rozhen.org }{narsenov@nao-rozhen.org }}\\
  \inst{1}\email{\href{mailto:peijin@nao-rozhen.org}{peijin@nao-rozhen.org}}}
\abstract{
Solar energetic particles are mainly protons and originate from the Sun during solar flares or coronal shock waves. Forecasting the Solar Energetic Protons (SEP) flux is critical for several operational sectors, such as communication and navigation systems, space exploration missions, and aviation flights, as the hazardous radiation may endanger astronauts’, aviation crew and passengers’ health, the delicate electronic components of satellites, space stations, and ground power stations. Therefore, the prediction of the SEP flux is of high importance to our lives and may help mitigate the negative impacts of one of the serious space weather transient phenomena on the near-Earth space environment. Numerous SEP prediction models are being developed with a variety of approaches, such as empirical models, probabilistic models, physics-based models, and AI-based models. In this work, we use the bi-directional long short-term memory (BiLSTM) neural network model architecture to train SEP forecasting models for 3 standard integral GOES channels ($>$10 MeV, $>$30 MeV, $>$60 MeV) with 3 forecast windows (1-day, 2-day, and 3-day ahead) based on daily data obtained from the OMNIWeb database from 1976 to 2019. As the SEP variability is modulated by the solar cycle, we select input parameters that capture the short-term, typically within a span of a few hours, and long-term, typically spanning several days, fluctuations in solar activity. We take the F10.7 index, the sunspot number, the time series of logarithm of the x-ray flux, the solar wind speed, and the average strength of the interplanetary magnetic field as input parameters to our model. The results are validated with an out-of-sample testing set and benchmarked with other types of models.
}


\maketitle
\section{Introduction}
\label{S_intro}
Solar Energetic Protons (SEP) are high-energy particles that are believed to be originated from the acceleration of particles in the solar corona during coronal mass ejections (CMEs) and solar flares \citep{aschwanden_2002, klein_2017, lin_2005, lin_2011, kahler_2017}. They are typically characterized by their high energy levels - with some particles having energies in the relativistic GeV/nucleon range - and their ability to penetrate through spacecraft shielding, causing radiation damage \citep{reames_2013, desai_2016}. The fluence and energy spectrum of SEP are influenced by several factors, including the strength of the solar flare or CME that produced them, and the conditions of the interplanetary environment \citep{kahler_1984, kahler_1987, debrunner_1988, miteva2013, trottet2015, dierckxsens2015, le2017, gopalswamy2017}.
SEP exhibit a strong association with the solar cycle, with the frequency and flux of SEP events peaking during the maximum phase of the solar cycle \citep{reames_2013}. This is thought to be due to the increased activity of the Sun during this phase, which leads to more frequent and powerful flares and CMEs. Previous studies have shown a relationship between the occurrence frequency of SEP and the sunspot number (SN; \citeauthor{nymmik_2007}, \citeyear{nymmik_2007}; \citeauthor{richardson2016}, \citeyear{richardson2016}). However, the exact relationship between the solar cycle and SEP is complex and not fully understood. Hence, more work is needed to better understand this connection, as previous studies have reported intense SEP events during relatively weak solar activity \citep{cohen_2018, ramstad_2018}. 

SEP have been a subject of interest and research in heliophysics for decades. It is hypothesized that shock waves generated in the corona can lead to an early acceleration of particles. However, SEP have sufficient energy to propagate themselves by \textit{surfing} the interplanetary magnetic fields (IMF), and therefore, the expanding CME is not necessary for their transport \citep{reames_2000, kota_2005, kozarev_2019, kozarev_2022}. While this theory has gained acceptance, there is an ongoing debate among scientists over the specific mechanisms and conditions responsible for SEP production and acceleration.
The creation, acceleration, and transport mechanisms of SEP are complex and involve a combination of magnetic reconnection, shock acceleration, and wave-particle interactions \citep{li_2003, li_2012, ng_2012}. The specific mechanisms responsible for SEP production and acceleration can vary depending on the type and strength of the solar event that triggered them. Further research is imperative to better understand the processes involved in the production and transport of SEP in the heliosphere. This will facilitate the development of more precise models that assist in minimizing the impact of SEP on astronauts and space-based assets. 

Several models are available, or under development, for forecasting SEP, which use diverse approaches and serve different objectives. These models comprise computationally complex physics-based models, quick and simple empirical models, Machine Learning (ML)-based models, and hybrid models that combine different approaches and produce different types of outputs, including deterministic, probabilistic, categorical, and binary. Deterministic models always generate the same output without any randomness or stochastic components, such as predicting the SEP flux at a specific moment or the arrival time of SEP. On the other hand, probabilistic models provide a probability value that reflects the likelihood of an SEP event occurring. However, replicating SEP fluxes at a specific time is still a significant challenge for current models.

An excellent review on SEP models and predictive efforts was recently published by \citet{whitman22}, which summarizes the majority of the existing models.
For instance, \citet{papaioannou_2022} introduced the Probabilistic Solar Particle Event Forecasting (PROSPER) model, which is incorporated into the Advanced Solar Particle Event Casting System (ASPECS)\footnote{\url{http://phobos-srv.space.noa.gr/}}. The PROSPER model utilizes a Bayesian approach and data-driven methodology to probabilistically predict SEP events for 3 integral energy channels $>$10, $>$30, and $>$100 MeV. The model's validation results indicate that the solar flare and CME modules have hit rates of 90\% and 100\%, respectively, while the combined flare and CME module has a hit rate of 100\%.
\citet{bruno_2021} developed an empirical model to predict the peak intensity and spectra of SEP at 1 AU between 10 and 130 MeV, using data from multiple spacecraft. The model is tested on 20 SEP events and shows good agreement with observed values. The spatial distribution of SEP intensities was reconstructed successfully, and they found a correlation between SEP intensities and CME speed.
\citet{hu_2017} extended the Particle Acceleration and Transport in the Heliosphere (PATH) model to study particle acceleration and transport at CME-driven shocks. They showed that the model can be used to obtain simultaneous calculations of SEP characteristics such as time-intensity profiles, instantaneous particle spectra, and particle pitch angle distributions at multiple heliospheric locations. Overall, results resemble closely those observed in situ near the Earth but also yield results at other places of interest, such as Mars, making it of particular interest to Mars missions.
SPREAdFAST \citep{kozarev_2017, kozarev_2022} is a physics-based, data-driven framework that utilizes EUV observations and models to simulate SEP fluxes at 1 AU and to estimate energetic particle acceleration and transport to various locations in the inner heliosphere. It generates time-dependent histograms and movies distributing them through an online catalog. The accuracy and efficiency of the model were encouraging, but the highest energy fluxes showed disagreement with in situ observations by the SOHO/ERNE instrument. However, the framework has great potential for space weather science and forecasting.

In \citet{aminalragia_2021}, they used neural networks to provide probabilities for the occurrence of SEP based on soft X-rays data from 1988 to 2013. They obtained $>$85\% for correct SEP occurrence predictions and $>$92\% for correct no-SEP predictions.
\citet{lavasa_2021} described a consistent approach to making a binary prediction of SEP events using ML and conventional statistical techniques. The study evaluated various ML models and concluded that random forests could be the best approach for an optimal sample comprising both flares and CMEs. The most important features for identifying SEP were found to be the CME speed, width, and flare soft X-ray fluence.
\citet{kasapis_2022} employed ML techniques to anticipate the occurrence of a SEP event in an active region that generates flares. They utilized the Space-Weather MDI Active Region Patches (SMARP) dataset, which comprises observations of solar magnetograms between June 1996 and August 2010. The SMARP dataset had a success rate of 72\% in accurately predicting whether an active region that produces a flare would result in a SEP event. Moreover, it provided a competitive lead time of 55.3 min in forecasting SEP events.

\citet{engell_2017} introduced the Space Radiation Intelligence System (SPRINTS), a technology that uses pre- and post-event data to forecast solar-driven events such as SEP. It integrates automatic detections and ML to produce forecasts. Results show that SPRINTS can predict SEP with an 56\% probability of detection and 34\% false alarm rate.
Nevertheless, the HESPERIA REleASE tools provide real-time predictions of the proton flux at L1 by using near-relativistic electrons as a warning for the later arrival of protons and have been set to operation\citep{malandraki_2018}. Historical data analysis indicates high prediction accuracy, with a low false alarm rate of approximately 30\% and a high probability of detection of 63\% \citep{malandraki_2018}.

Forecasting SEP is a critical task that serves operational needs and provides insight into the broader field of space weather science and heliophysics. As emphasized in previous works, a high precision forecasting model is urgently required to predict SEP flux within a period of time, given the risks associated with these events. This highlights the critical requirement for a dependable forecasting system that can mitigate the risks associated with SEP.

Scientists have been using physics-based and empirical models for decades to forecast SEP. However, these models have certain limitations. Physics-based models require accurate input data and underlying physical assumptions. In addition, the complexity of the physics involved and incorrect parameters may introduce uncertainties that can lead to inaccurate predictions.
On the other hand, empirical models rely on historical data to make predictions.
While they can be accurate sometimes, they may be unable to account for changes in physical conditions related to the acceleration and propagation of SEP, which can influence prediction accuracy.
ML models, however, provide a different approach to SEP forecasting. These models can analyze vast amounts of data, learning patterns from the data that are used, and connections that may not be obvious to experts. Additionally, ML models can adapt to changes in underlying physical conditions, resulting in more accurate predictions as more data is collected; they also provide relatively rapid forecasts, which allows for incorporation into a real-time forecasting workflow.

In the upcoming sections, we will explore the limitations in accuracy that arise from dealing with an imbalanced dataset and low-resolution data. Specifically, the presence of intrinsic outliers in the time series data pertaining to SEP flux poses a significant challenge in modeling. These outliers correspond to occurrences of SEP events and, consequently, have an impact on the accuracy of predictions. Notably, they often lead to an underestimation of the SEP fluxes, primarily due to the predominance of relatively low values throughout the majority of the time interval.

In this paper, we present advanced deep learning models to forecast the daily integral flux of SEP over a 3-day forecasting window by using bi-directional long short-term memory (BiLSTM) neural networks, for 3 energy channels ($>$10, $>$30, and $>$60 MeV).
Our models can forecast the time-dependent development of SEP events in different energy domains, which can be used to model the space radiation profiles using frameworks such as BRYNTRN \cite{wilson_1988} and GEANT4 \citep{truscott_2000}.
We describe the data selection and pre-processing in Section~\ref{S_data_prep}. We present an overview on the analysis methods and the models implemented in Section~\ref{S_methods}. Then we show the forecasting results in Section~\ref{S_results}. Finally the summary and implications are presented in Section~\ref{S_conclusion}.
\section{Data Preparation}
\label{S_data_prep}
In this section, we describe the physical quantities, the types of inputs and their sources, as well as the outputs we are forecasting.
Some of the technical terms used in this study are explained further in the appendices.

In order to capture the variability of solar activity, which modulates the SEP flux, we selected input physical quantities that describe both the interplanetary medium and solar activity. These input features can be categorized into two groups: remote signatures and in-situ measurements.
The remote signatures consist of the F10.7 index, as well as the long-wavelength ($X_L$) and short-wavelength ($X_S$) x-ray fluxes. The F10.7 index represents the flux of solar radio emission at a wavelength of 10.7 cm, measured in solar flux units (sfu). To obtain the x-ray fluxes, we utilized 1- and 5-minute averaged data from the Geostationary Operational Environmental Satellite (GOES) database\footnote{\url{https://satdat.ngdc.noaa.gov/sem/goes/data/avg/}}, specifically at long wavelengths (1 - 8 \AA) and short wavelengths (0.5 - 4.0 \AA).

The in-situ measurements encompass the near-Earth solar wind magnetic field and plasma parameters. These include the solar wind speed (in km s$^{-1}$), average interplanetary magnetic field (IMF) strength (in nT), and the integral SEP fluxes at three energy channels: $>$10, $>$30, and $>$60 MeV, which correspond to the GOES channels (in 1/cm$^2$ sec ster). These SEP fluxes were obtained from multiple spacecraft stationed at the first Lagrange point (L1) throughout the study period.
In particular, the IMF and plasma data in the OMNI database are obtained from the IMP, Wind, and ACE missions, while the energetic particle fluxes are obtained from the IMP and GOES spacecraft\footnote{\url{https://omniweb.gsfc.nasa.gov/html/ow_data.html}}.

To ensure a comprehensive dataset, we acquired hourly-averaged data covering a timeframe from December 1976 to July 2019, which spans the past four solar cycles. These data were sourced from the Space Physics Data Facility (SPDF) OMNIWeb database\footnote{\url{https://omniweb.gsfc.nasa.gov}}, hosted by the Goddard Space Flight Center. This database provides a wealth of information, including integral proton fluxes, as well as an extensive range of solar wind plasma and magnetic field parameters.
Lastly, the daily data on sunspot numbers were obtained from the Sunspot Index and Long-term Solar Observations (SILSO) archive\footnote{\url{https://www.sidc.be/silso/home}}, maintained by the World Data Center.

\begin{figure}[h!]
    \centerline{\includegraphics[width=0.9\textwidth]{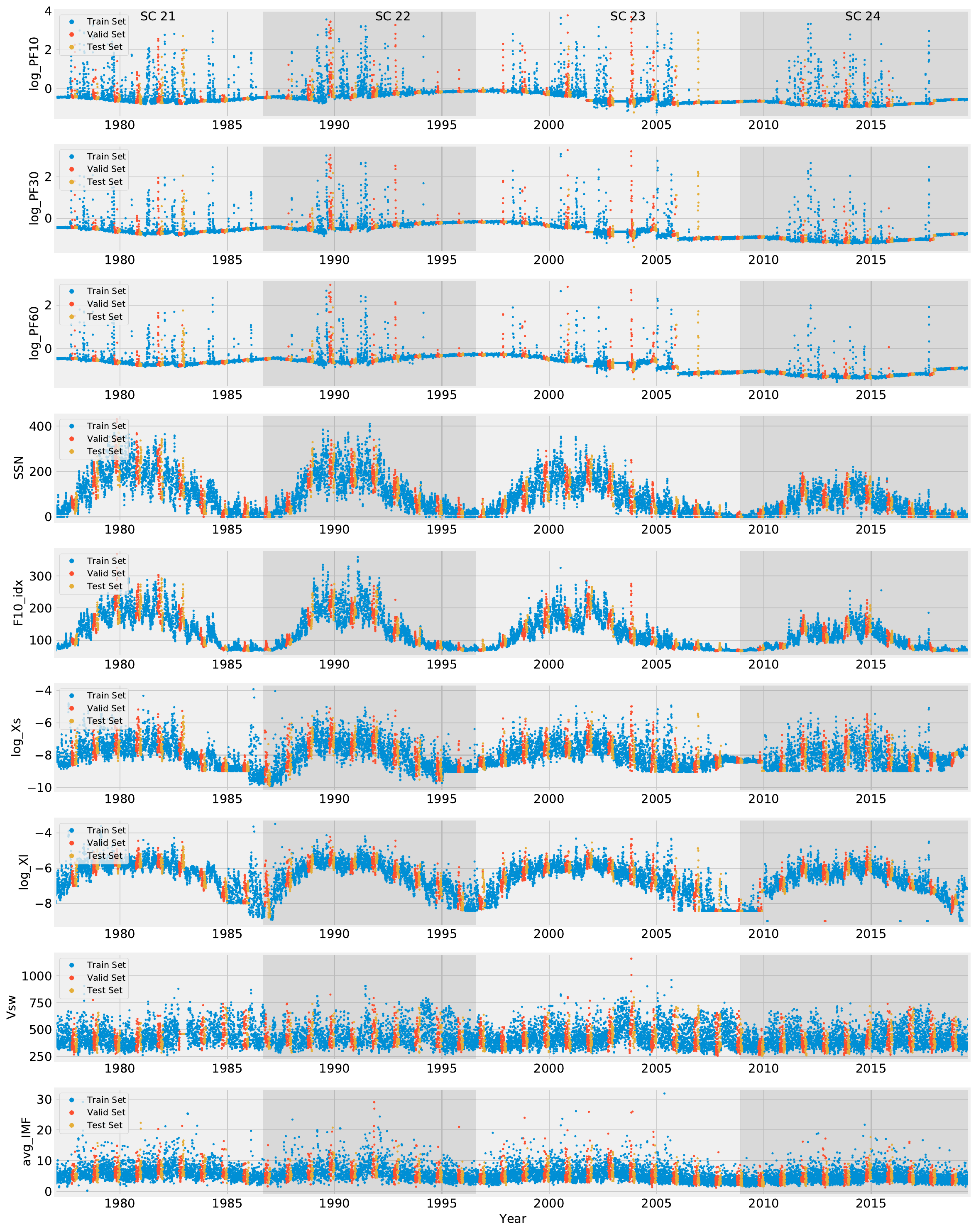}}
    \caption{Data splitting for all input features, showing the training, validation, and testing sets. Daily data from 1976-12-25 00:00 to 2019-07-30 00:00. The gray shading labels the solar cycles from SC21 to SC24.}
\label{fig_allFeatures}
\end{figure}

Figure~\ref{fig_allFeatures} shows a plot for the timeseries data of all features.
The top 3 panels are the logarithms of the SEP integral flux at the 3 energy channels (log\_PF10, log\_PF30, and log\_PF60), then the sunspot number, the F10.7 index (F10\_idx), the logarithms of the x-ray fluxes (log\_Xs and log\_Xl), the solar wind speed (Vsw), and the average magnitude of the IMF (avg\_IMF).
Throughout this paper, we adopt the convention that "log" refers to the common logarithm with a base of 10.
The gray shades refer to the timespan of solar cycles.
The blue, orange, and gold colors refer to the training, validation, and test sets, respectively. The data split method will be explained shortly.

Since the input SEP data have been compiled from various spacecraft, it may have artifacts even after processing. In particular, there are occasional jumps in the background level. There are also several day-long gaps in the OMNI solar wind parameters from the early 1980s to mid-1990s where only IMP 8 data are available and this spacecraft spent part of each orbit in the magnetosphere. We are reasonably confident that these issues do not influence the overall analysis significantly.

In deep learning applications, the dataset is split into 3 sets; namely the training set, the validation set, and the test set. The training set is usually the largest chunk of data that is used to fit the model. The validation set is a smaller chunk of data used to fine-tune the model and evaluate its accuracy to ensure it is unbiased. The test set is the out-of-sample data exclusively used to assess the final model when performing on unseen data \citep{ripley_2007}.

\begin{figure}[htp]
    \centerline{\includegraphics[width=\textwidth]{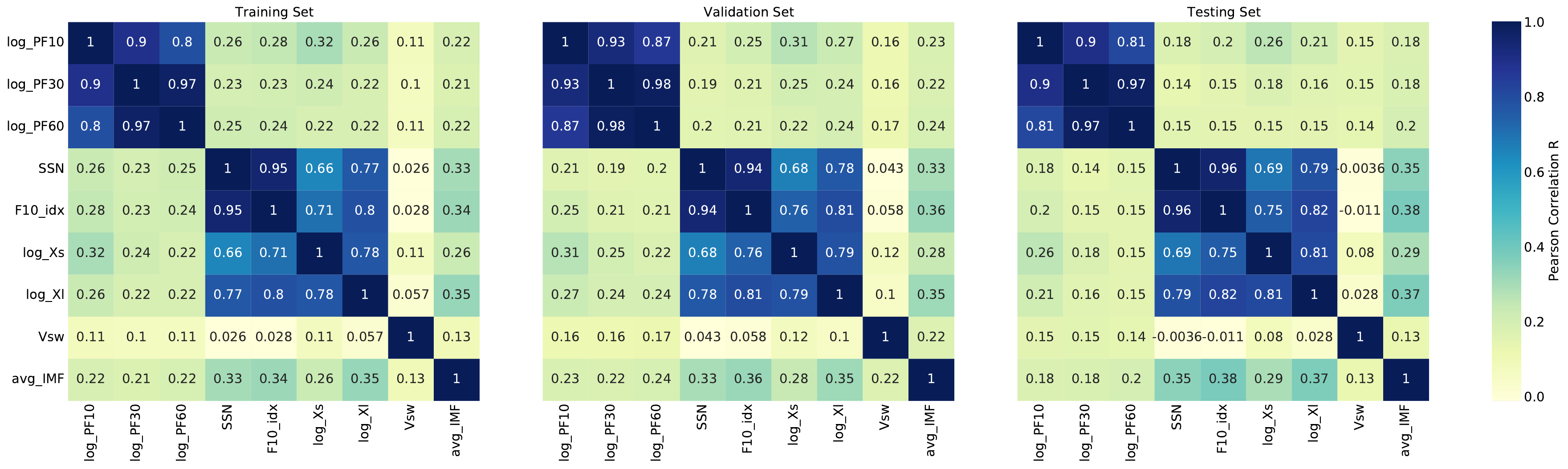}}
    \caption{Correlation matrices show the correlation between the features in the training, validation, and test sets.}
\label{fig_dataCorr}
\end{figure}

After inspecting the correlation between the solar wind indices and the SEP integral fluxes in the OMNIWeb database, we chose the top-correlated features with the SEP flux. The correlations were made between the SEP fluxes and the individual parameters. Hence we took only timeseries of logarithms of the protons' integral flux at 3 energy channels ($>$10, $>$30, and $>$60 MeV), the timeseries of logarithm of the X-ray fluxes, the F10.7 index, the sunspot number, the solar wind speed, and the average strength of the IMF as input parameters to our model.
The log of the SEP flux was used across the whole study.
The correlation matrices for the training, validation, and test sets are shown in Figure~\ref{fig_dataCorr}.
The X-ray and proton fluxes were converted into the logarithmic form because it was more convenient than the original form of data since the time series data were mostly quiet and had numerous sharp spikes, which correspond to solar events.
Based on a previous experience with NNs \citep{mnedal_2019}, we found that training separate models for each target (output) feature can lead to better results. This is because a dedicated model for each  output feature can more easily learn the interrelationships between input features and make more accurate predictions. Therefore, in our current study, we trained 3 separate models, each one targeting the logarithm of the protons integral flux at a specific energy channel.

In order to ensure consistency across all features, all durations of the time series data of the physical quantities were matched to be within the same time range. Subsequently, the dataset was resampled to obtain daily averaged data, resulting in a significant reduction of the dataset size by a factor of 24. This reduction facilitated expeditious training and yielded prompt results.

There were missing data values in the original dataset; for the $B_{avg}$ ($\sim$10.7\%), $V_{sw}$ ($\sim$10.5\%), F10.7-index ($\sim$0.08\%), short-band x-ray flux ($\sim$8\%), long-band x-ray flux ($\sim$9.8\%), and proton fluxes ($\sim$4.3\%). The data gaps were linearly interpolated.

In timeseries forecasting, it is a common practise to take a continuous set of data points from the main dataset to be the validation set and another smaller chunk of data to be the test set, for instance in \citet{pala_2019, benson20, zhang_2022, zhu_2022}. 
From our experiments, we got descent results when we applied the same data split method, but the results were a bit biased toward the end of the solar cycle 24 and the testing set was biased towards a quiet period. So, we adopted the 9-2-1 strategy, that is taking from each year 9 months to be added in the training set, 2 months to be added in the validation set, and 1 month to be added in the test set. This is applied over the $\sim$43 years of data (Fig.~\ref{fig_allFeatures}), which yields 74.29\% of the data for the training set, 16.2\% for the validation set, and 9.51\% for the testing set. By doing so, we eliminated the need to do cross-validation and hence, made the training more efficient.
It is worth to mention that the timeseries data must not be shuffled as that will break temporal and logical order of measurements, which must be maintained.

\section{Methods}
\label{S_methods}
In this section, we introduce the data analysis methods used in this work. We start with explaining the model selection phase, followed by a discussion of the bidirectional long-short-term memory (BiLSTM) neural network architecture. The technical terminologies are described in the appendices.

\subsection{The Bi-LSTM Model}
\begin{figure}[htp]
    \centerline{\includegraphics[width=0.6\textwidth]{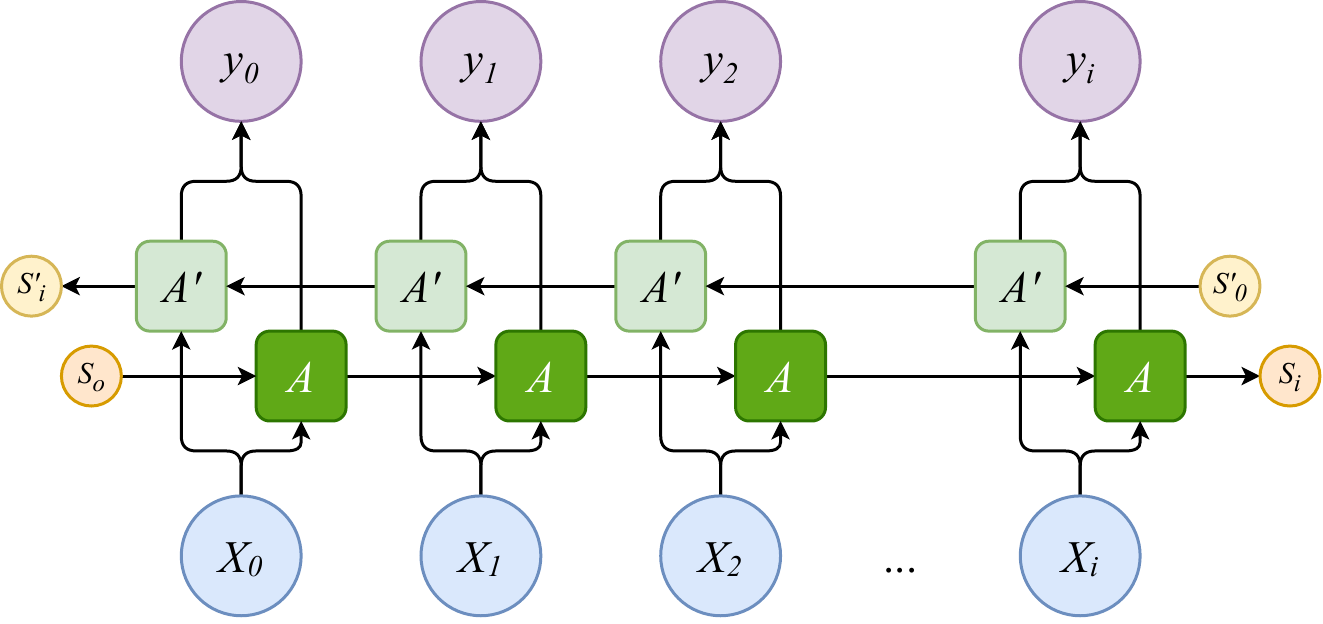}}
    \caption{Architecture of a single BiLSTM layer. The blue circles at the bottom labeled by \textit{($x_0$, $x_1$, $x_2$, ..., $x_i$)} are the input data values at multiple time steps. The purple circles, on the other hand, are the output data values at multiple time steps labeled by \textit{($y_0$, $y_1$, $y_2$, ..., $y_i$)}. The dark green and light green boxes are the activation units of the forward layer and the backward layer, respectively. The orange and yellow circles are the hidden states at the forward layer and the backward layer, respectively. Both the forward and backward layers composes a single hidden BiLSTM layer. The figure is adopted from \citet{olah_2015}}
\label{fig_model}
\end{figure}

Recurrent neural networks (RNNs) that support processing input sequences both forward and backward are known as Bidirectional Long Short-Term Memory (BiLSTM) neural networks \citep{schuster_1997}. Regular RNNs \citep{hochreiter_1997, kolen2001gradient} depend on the prior hidden state and the current input to determine the output at a given time. The output of a BiLSTM network, on the other hand, is dependent on the input at a given moment as well as the previous and future hidden states. As a result, the network is able to make predictions using contexts from the past as well as the future. Hence, accuracy is improving.
Each BiLSTM layer consists of two LSTM layers; a forward layer that processes the input sequences from the past to future, and a backward layer that processes the input sequences from the future to the past, as illustrated in Figure~\ref{fig_model}, to capture information from both past and future contexts. The output from each layer is concatenated and fed to the next layer, which can be another BiLSTM layer or a fully connected layer for final prediction.

BiLSTM networks are advantageous than traditional LSTM networks in a variety of aspects \citep{graves_2005, ihianle_2020, alharbi_2021}. First, as we demonstrate in this study, they are excellent for tasks like timeseries forecasting, as well as speech recognition and language translation \citep{wollmer2013, graves2014, sundermeyer2014, Huang2018, nammous2022} because they can capture long-term dependencies in the input sequence in both forward and backward directions. Second, unlike feedforward networks, BiLSTM networks do not demand fixed-length input sequences, thus being able to handle variable-length sequences better. Furthermore, by taking into account both past and future contexts, BiLSTM networks can handle noisy data.
However, BiLSTM networks are computationally more expensive than regular LSTM networks due to the need for processing the input sequence in both directions. They also have a higher number of parameters and require more training data to achieve good performance.

The final dataset has 7 features, including the target feature, from December 25$^{th}$ 1976 to July 30$^{th}$ 2019, with a total of 15,558 samples (number of days). The training set has 11,558 samples, the validation set has 2,520 samples, and the test set has 1,480 samples.

The input horizon of 270 steps (30 days × 9 months) was used.
A data batch size of 30 was used, which is the number of samples processed that result in one update to the model's weights (Appendix~\ref{bilstm_appendix}).
The model consists of 4 BiLSTM layers with 64 neurons each, and an output dense layer with 3 neurons, representing the output forecasting horizon.
The total number of trainable parameters is 333,699.
The number of training epochs was set to 50 because from experiments, the model stopped improving remarkably after almost 50 epochs. Thus, there was no need to waste time and computational resources to train the model for more than 50 epochs.

The \textit{ModelCheckpoint} callback function was used to register the model version with the minimal validation loss. 
The \textit{EarlyStopping} callback function was used to halt the model run when detecting overfitting, with a \textit{patience} parameter of 7. 
\textit{ReduceLROnPlateau} callback function was used to reduce the learning rate when the validation loss stops improving, with a \textit{patience} parameter of 5, a reduction factor of 0.1 and minimal learning rate of 1e$^{-6}$.
\subsection{Model Selection}
To determine the most suitable model for our objective and provide justifiable reasons, we conducted the following analysis.
First we examined the naive (persistence) model, which is very simplistic and assumes that the timeseries values will remain constant in the future. In other words, it assumes that the future value will be the same as the most recent historical value. That was the baseline. Next we examined the moving-average model, which calculates the future values based on the average value of historical data within a specific time widow. This gives a little bit lower error.

\begin{figure}[htp]
    \centerline{\includegraphics[width=0.5\textwidth]{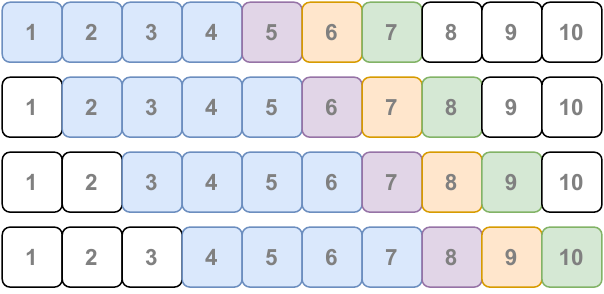}}
    \caption{Illustration of the sliding window technique for a sample of 10 timesteps, where each number denotes a distinct time step. As an example here, the input horizon (blue color) length is 4 timesteps and the output horizon length is 3 timesteps. The input window slides 1 time step at a time across the entire data sequence to generate 4 distinct input and forecast horizon pairs. The purple, orange, and green colors of the output horizon represent 1-day, 2-day, and 3-day ahead forecasting, respectively. The timesteps of 1-day ahead forecasting across the data sequences are then concatenated into a single timeseries list that is called 1-day ahead prediction. The same for 2-day and 3-day ahead.}
\label{fig_slide_window}
\end{figure}

After that, we went towards the machine learning (ML)-based models. For all the ML models, we chose the Adaptive moment estimation (Adam) optimizer \citep{kingma_2015} as the optimization algorithm due to its minimal memory requirements and high computational efficiency as it is well-suited for applications that involve large number of parameters or large datasets. As a rule of thumb, we set the optimizer’s learning rate to be 0.001 as it is usually recommended \citep{zhang_2022}.

In order to prepare the data in a readable format to the ML models, we created a windowed dataset with an input horizon of 365 steps representing 1 year of data and an output horizon of 3 steps representing the forecast window of three days. We call this windowing method as Multi-Input Multiple Output (MIMO) strategy, in which the entire output sequence is predicted in one shot. The MIMO strategy adopts the sliding window method that was mentioned in \citet{benson20} in which each sequence is shifted by one step with respect to the previous sequence until reaching the end of the available data (Fig.~\ref{fig_slide_window}).
This approach minimized the imbalance of active days, with high SEP fluxes, and quiet days.

\begin{figure}[htp]
    \centerline{\includegraphics[width=\textwidth]{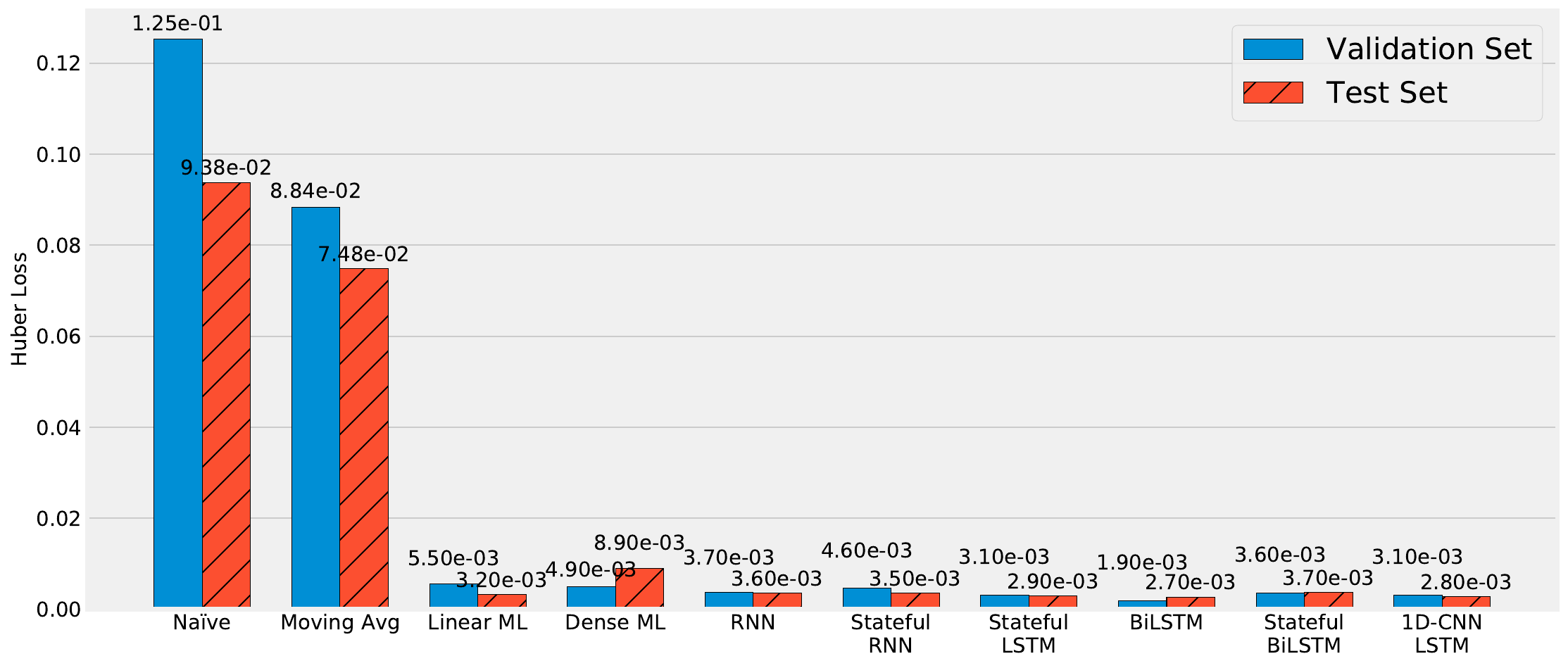}}
    \caption{Benchmarking of 10 models, shows the Huber loss for the validation and test sets.}
\label{fig_benchmark}
\end{figure}

After experiments with different loss functions and evaluate their performance on our dataset, we chose the Huber function~\ref{eq_huber} as the loss function and the Mean Absolute Error (MAE) is used as the metric function to monitor the model performance.
We used the Huber function because it is robust and combines the advantages of both Mean Squared Error (MSE) and MAE loss functions. It is less sensitive to outliers than MSE, while still being differentiable and providing gradients, unlike MAE. Since our data is noisy and contains outliers that may negatively impact the model's performance, the Huber loss function is a good choice.

We examined various neural network models to determine the optimal architecture for our task.
Initially, we started with a simple linear model comprising of a single layer with a single neuron. However, this model did not yield satisfactory results.
We then explored a dense ML model consisting of two hidden layers, each with 32 neurons and a \textit{RelU} activation function. Next, we experimented with a simple RNN model with the same number of hidden layers and neurons.
To find the optimal learning rate, we utilized the \textit{LearningRateScheduler} callback function and discovered that a rate of 1.58e$^{-4}$ under the basic settings minimized the loss.
We proceeded to examine stateful versions of RNN, LSTM, and BiLSTM models with three hidden layers, each with 32 neurons and a learning rate of 1.58e$^{-4}$.
In addition, we explored a hybrid model that consisted of a 1-dimensional convolutional layer with 32 filters, a kernel size of 5, and a \textit{RelU} activation function. We combined this with a two-hidden layer LSTM network with 32 neurons each and a learning rate of 1.58e$^{-4}$. We experimented with \textit{Dropout} layers but did not observe any significant improvement in the results.
Finally, we evaluated a BiLSTM model with five hidden layers, 64 neurons each, and a learning rate of 0.001.
Based on the evaluation of all the models on both the validation and test sets (Fig.~\ref{fig_benchmark} and Table~\ref{table_models_config}), we selected the BiLSTM model for further refinement. More details on the final model architecture and hyperparameters are explained in the Appendix~\ref{config_appendix}.
Figure~\ref{fig_benchmark} presents a comparative analysis of the Huber loss within the validation and testing sets across the ten aforementioned models.
We used several evaluation measures to assess our models since each metric provides valuable insights into the accuracy and performance of the forecasts (Appendix~\ref{eval_appendix}), helping to identify areas for improvement and adjust the forecasting models accordingly.
\section{Results and Discussion}
\label{S_results}
The benchmarking in Figure~\ref{fig_benchmark} showed that, in general, the ML-based methods were not much different. On the other hand, the persistence model and moving average model resulted in the highest errors compared with the ML-based models, and their results were close to some extent. 
As we see, the BiLSTM model performed the best over both the validation and test sets compared with the other models.

\begin{figure}[htp]
    \centerline{\includegraphics[width=\textwidth]{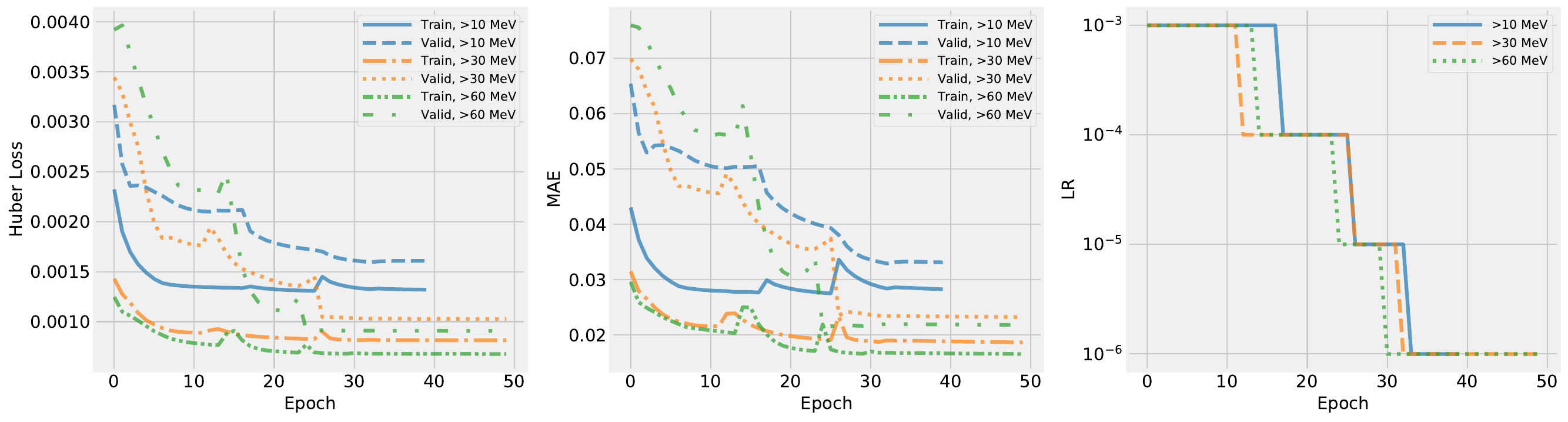}}
    \caption{\textit{Left Panel} - The Huber loss vs. the number of training epochs for the BiLSTM model for the validation and test sets, for the 3 energy channels. \textit{Middle Panel} - The mean absolute error (MAE); the model's metric vs. the number of training epochs. \textit{Right Panel} - Shows how the learning rate of the Adam optimizer changes over the number of epochs.}
\label{fig_lossCurve}
\end{figure}

We developed and trained 3 BiLSTM models to forecast the integral flux of SEP, one model per energy channels. After the training was completed, we evaluated the performance of the models from the loss curve (Fig.~\ref{fig_lossCurve}) using the Huber loss (the left panel) and the metric MAE (the middle panel). During the training, the learning rate was reduced multiple times via the \textit{LearningRateScheduler} callback function (the right panel).
The left panel quantifies the discrepancy between the model's predictions and the true values over time. It shows how the Huber loss function changes during the training iterations (Epochs) for the training and validation sets for the three energy channels so that each channel has one color.
The middle panel shows how the model's metric MAE changes with training epochs. It is used to evaluate the performance of the trained model by measuring the average absolute difference between the model's predictions and the true values, providing a single numerical value that indicates the model's error at a given epoch.
The right panel shows how the learning rate of the model's optimizer changes with epochs via the \textit{LearningRateScheduler} callback function, which changes the learning rate based on a predefined schedule to improve training efficiency and convergence.
The learning rate refers to the rate at which the model's parameters are updated during the training process.
We noticed that at the epochs where the learning rate has changed, there were bumps in the loss curves across all the energy channels, which is expected. This highlights the boundaries within which the learning rate yields better performance.

From experimentation, we found that the batch size and the optimizer learning rate are the most important hyperparameters that have a strong influence on the overall model's performance \citep{greff_2016}.
In addition, adding \textit{dropout} layers as well as varying the number of hidden layers and hidden neurons resulted in only marginal improvements to the final model performance, while substantially increasing training time and requiring greater computational resources.

\begin{figure}[htp]
    \centering
    \begin{subfigure}
         \centering
         \includegraphics[width=0.4\textwidth]{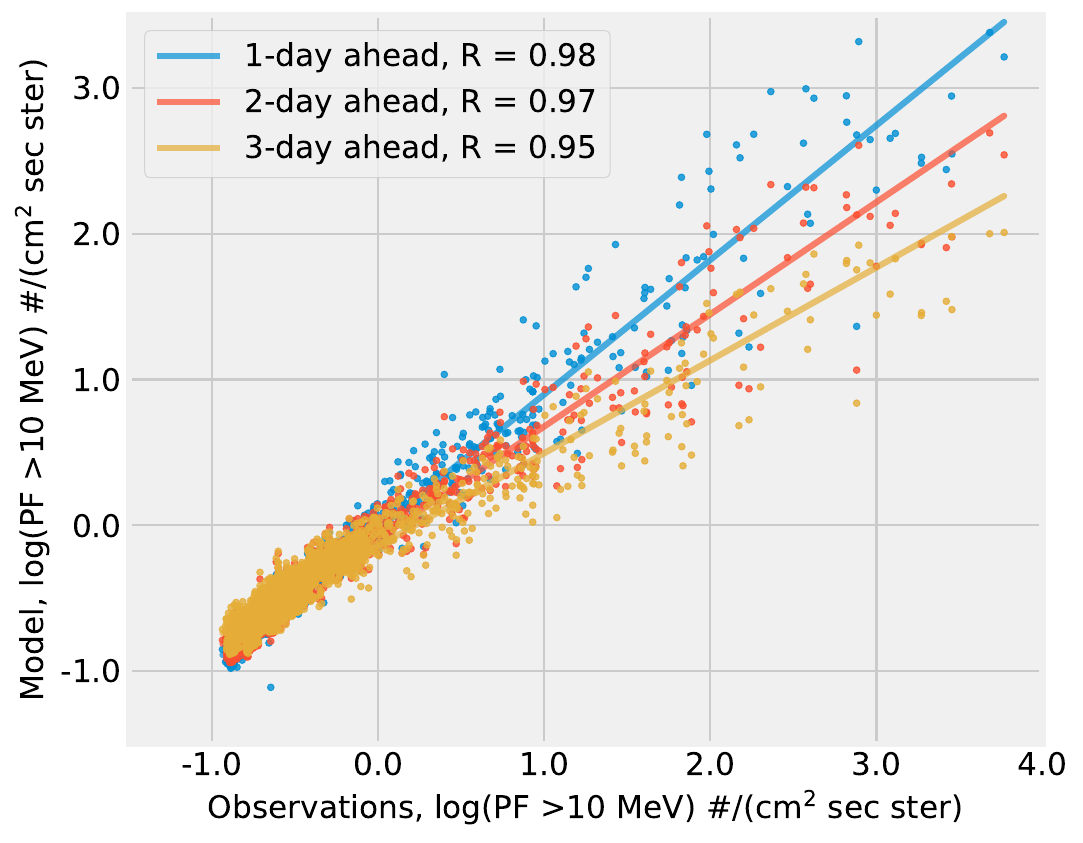}
    \end{subfigure}
    \begin{subfigure}
         \centering
         \includegraphics[width=0.4\textwidth]{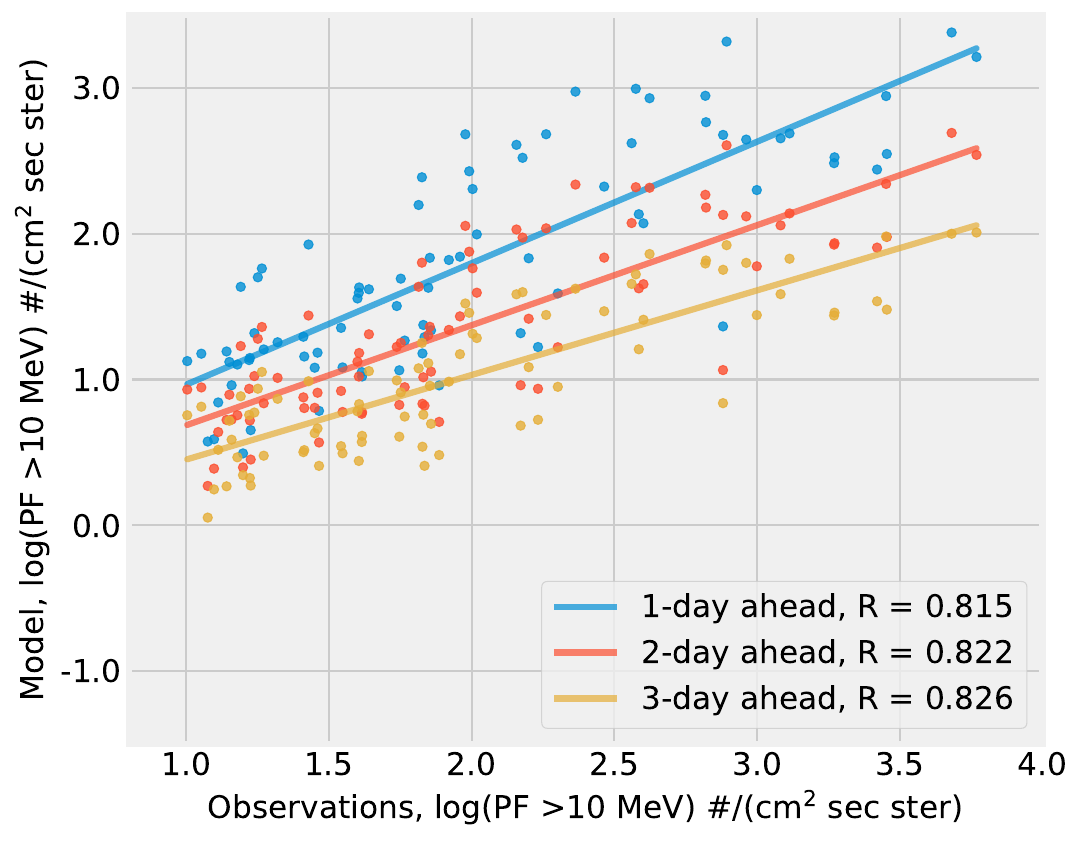}
    \end{subfigure}
    \begin{subfigure}
         \centering
         \includegraphics[width=0.4\textwidth]{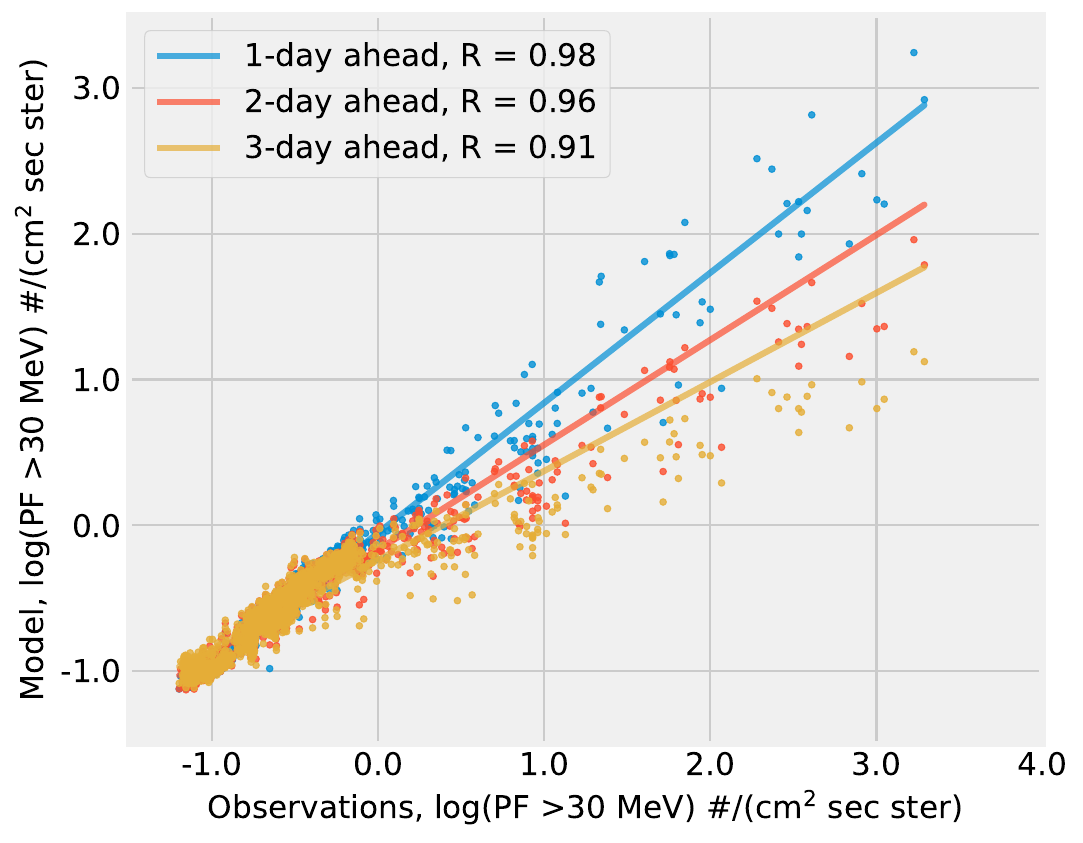}
    \end{subfigure}
    \begin{subfigure}
         \centering
         \includegraphics[width=0.4\textwidth]{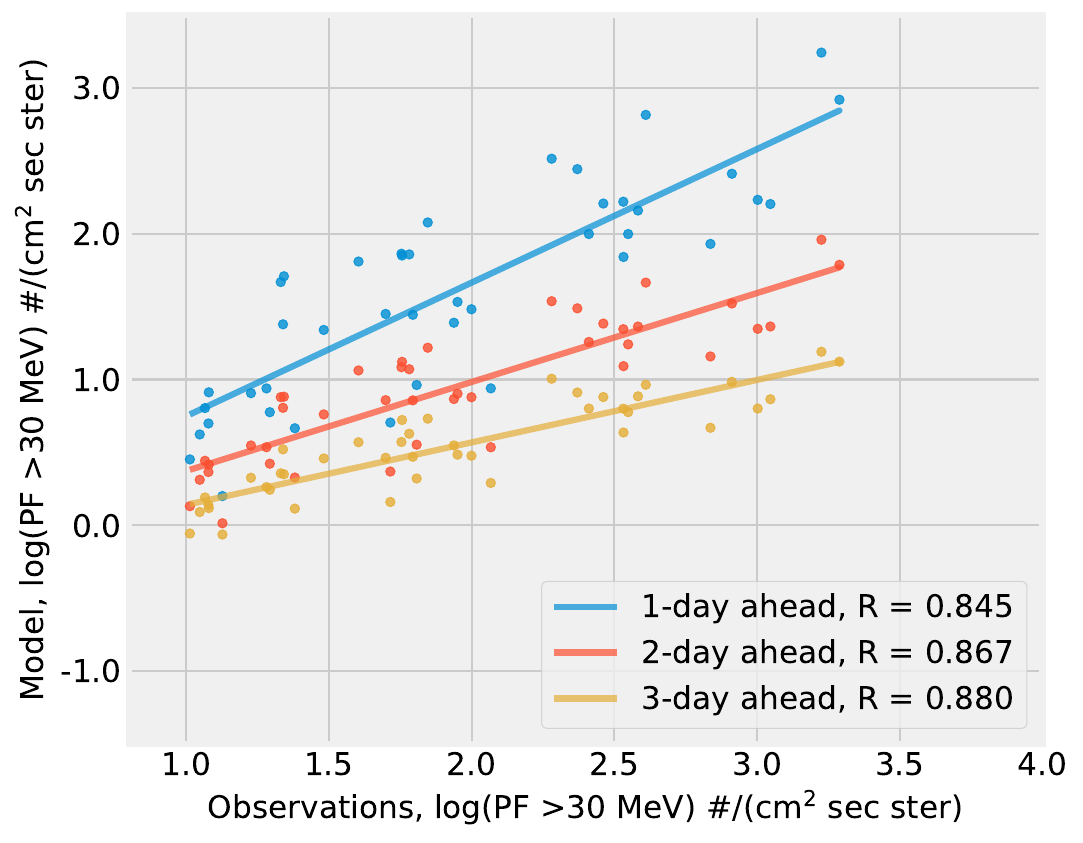}
    \end{subfigure}
    \begin{subfigure}
         \centering
         \includegraphics[width=0.4\textwidth]{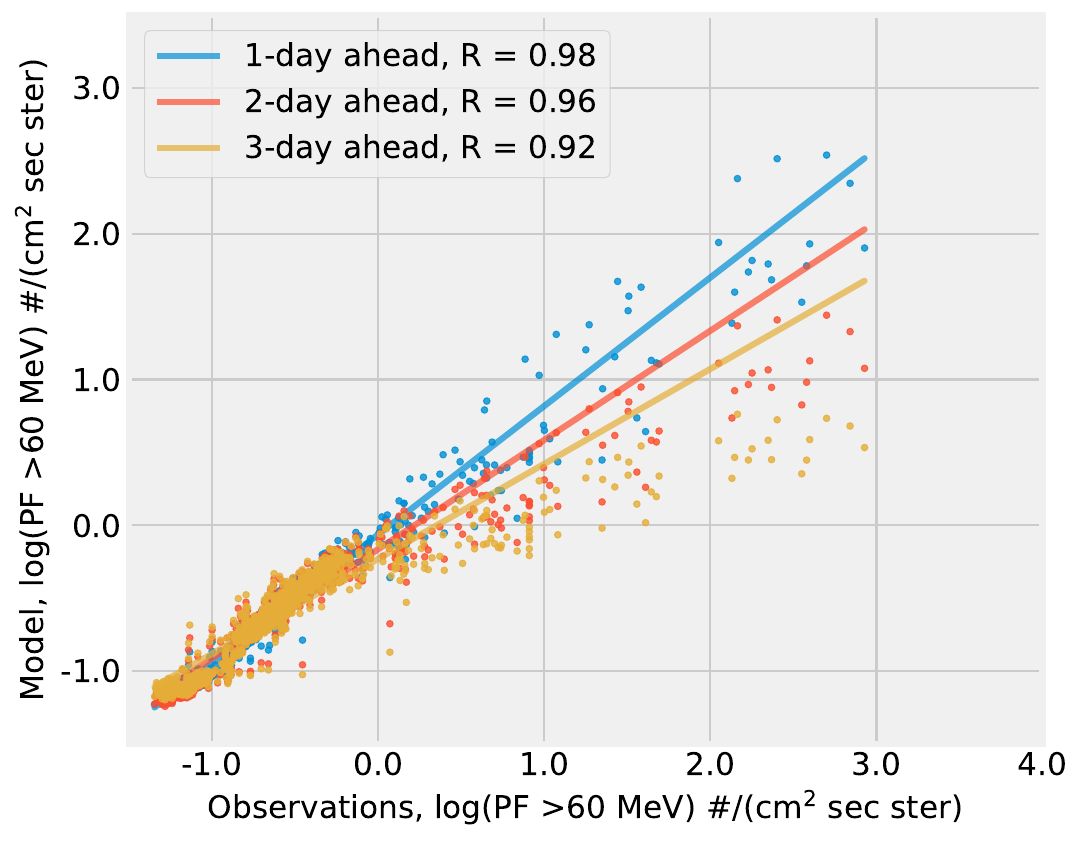}
    \end{subfigure}
    \begin{subfigure}
         \centering
         \includegraphics[width=0.4\textwidth]{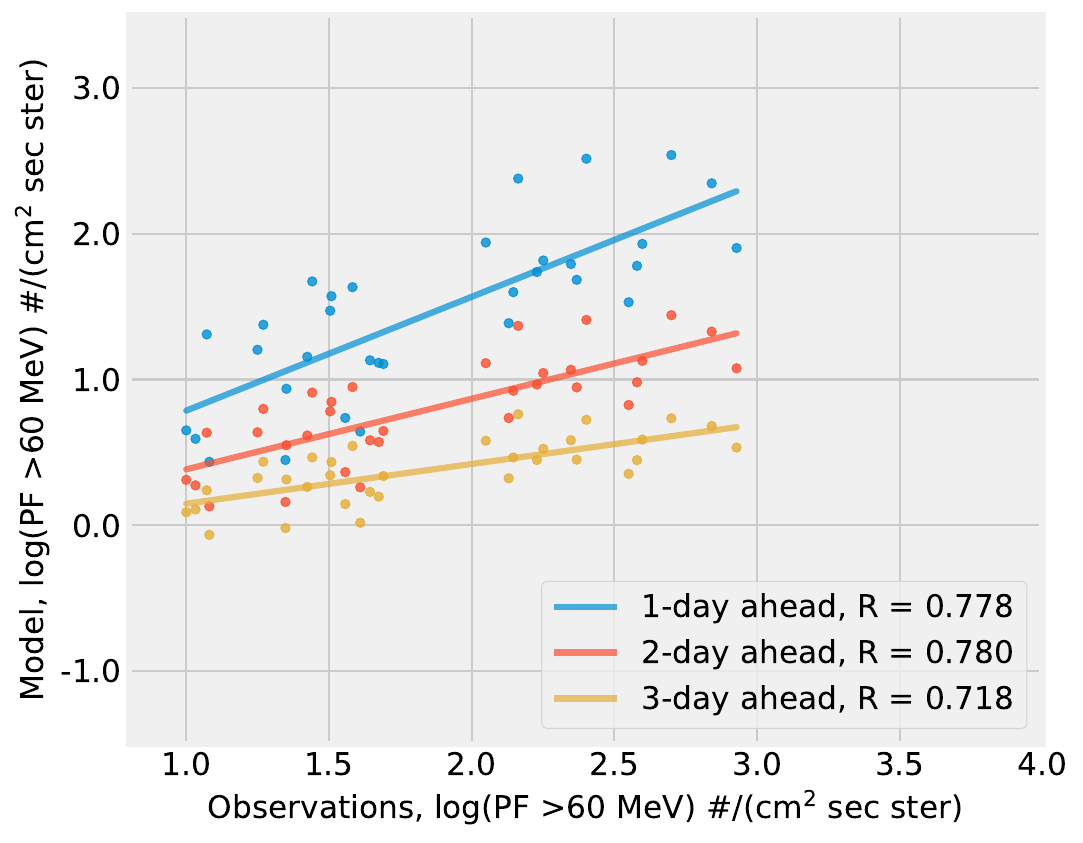}
    \end{subfigure}
\caption{Correlation between the model predictions and observations for 1-day, 2-day, and 3-day ahead for $>$10 MeV (top panel), $>$30 MeV (middle panel), and $>$60 MeV (bottom panel). The panels in the left column represent all the points of the validation set, those in the right column represent all the observations points with daily mean flux $\geq$10 pfu.}
\label{fig_model_vs_obs_valset}
\end{figure}

The term \textit{batch size} refers to the number of data sequences processed in one iteration during the training of a ML model \citep{goodfellow_2016}. Initially, a batch size of 64 was selected, however, we observed that the model produced better results when a batch size of 30 was used instead. This could be related to the Carrington rotation, which lasts for $\sim$27 days. There were $\sim$570 Carrington rotations between December 25$^{th}$ 1976 and July 30$^{th}$ 2019. Therefore, updating the model's weights after every Carrington rotation could be a reasonable choice for improving its performance.
Figure~\ref{fig_model_vs_obs_valset} shows how good the model predictions are (on the y-axis) compared with the observations of the validation set (on the x-axis). The blue, orange, and gold colors refer to 1-day, 2-day, and 3-day ahead predictions, respectively. The top panel is for the $>$10 MeV channel, the middle panel is for the $>$30 MeV channel, and the bottom panel is for the $>$60 MeV channel. The left column is for the entire validation set, while the right column is for the observations points $\geq$10 proton flux units (pfu). That is the threshold value of proton flux as measured by the National Oceanic and Atmospheric Administration (NOAA) GOES spacecraft to indicate severity of space weather events caused by SEP.

\begin{figure}[htp]
    \centering
    \begin{subfigure}
         \centering
         \includegraphics[width=0.4\textwidth]{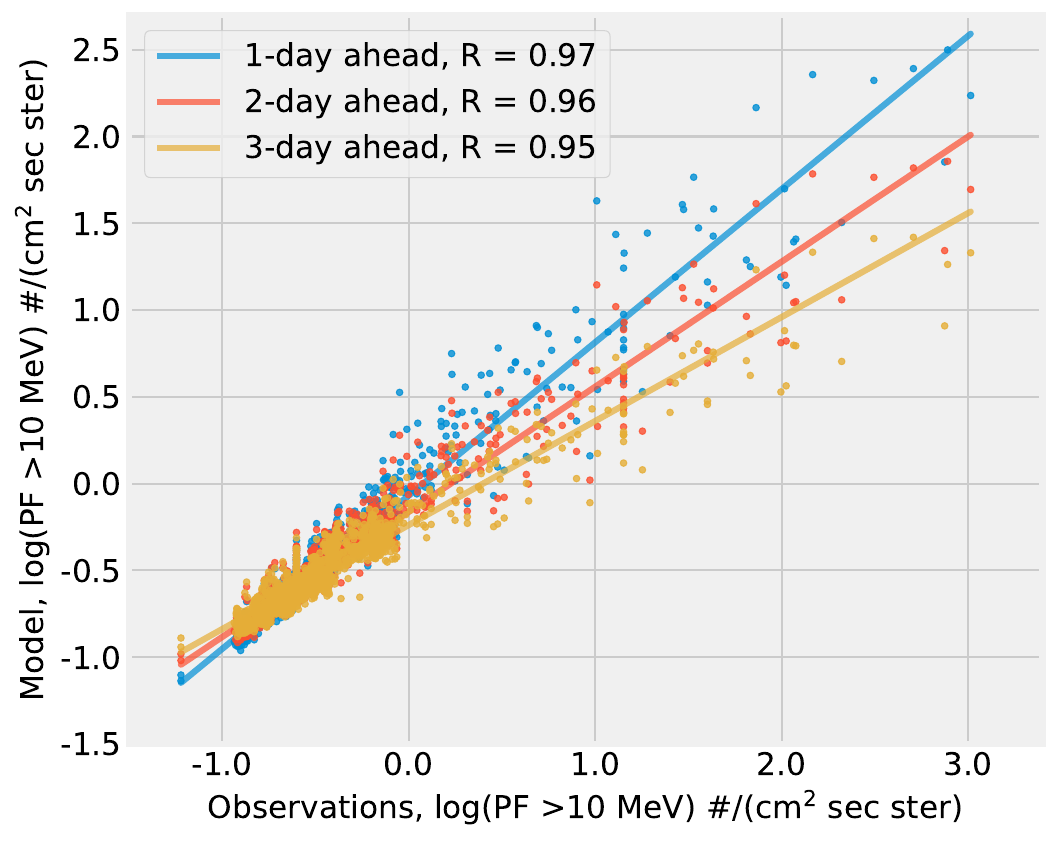}
    \end{subfigure}
    \begin{subfigure}
         \centering
         \includegraphics[width=0.4\textwidth]{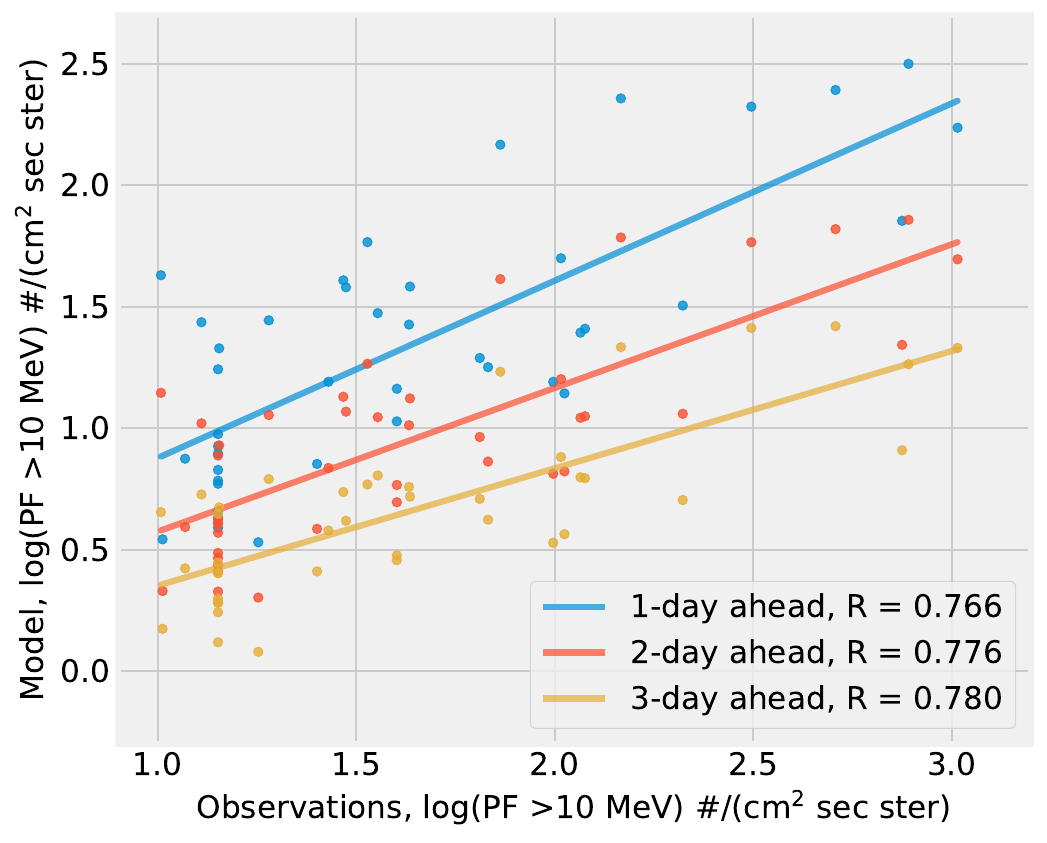}
    \end{subfigure}
    \begin{subfigure}
         \centering
         \includegraphics[width=0.4\textwidth]{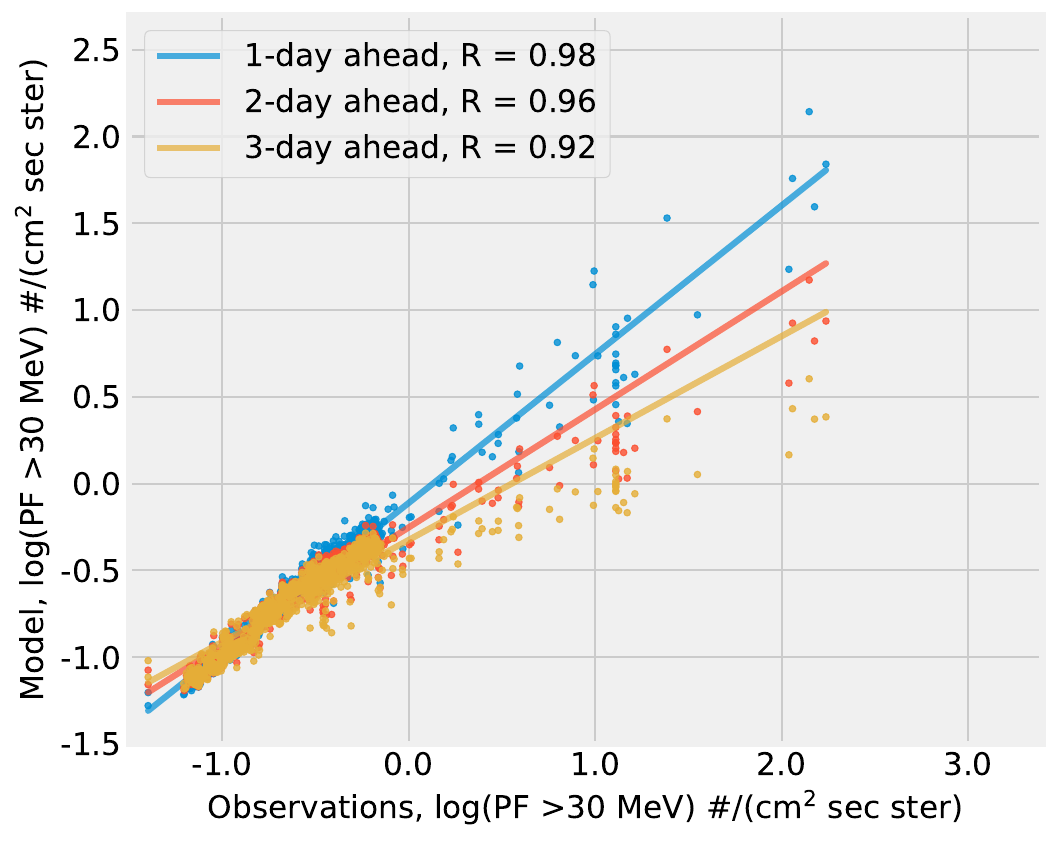}
    \end{subfigure}
    \begin{subfigure}
         \centering
         \includegraphics[width=0.4\textwidth]{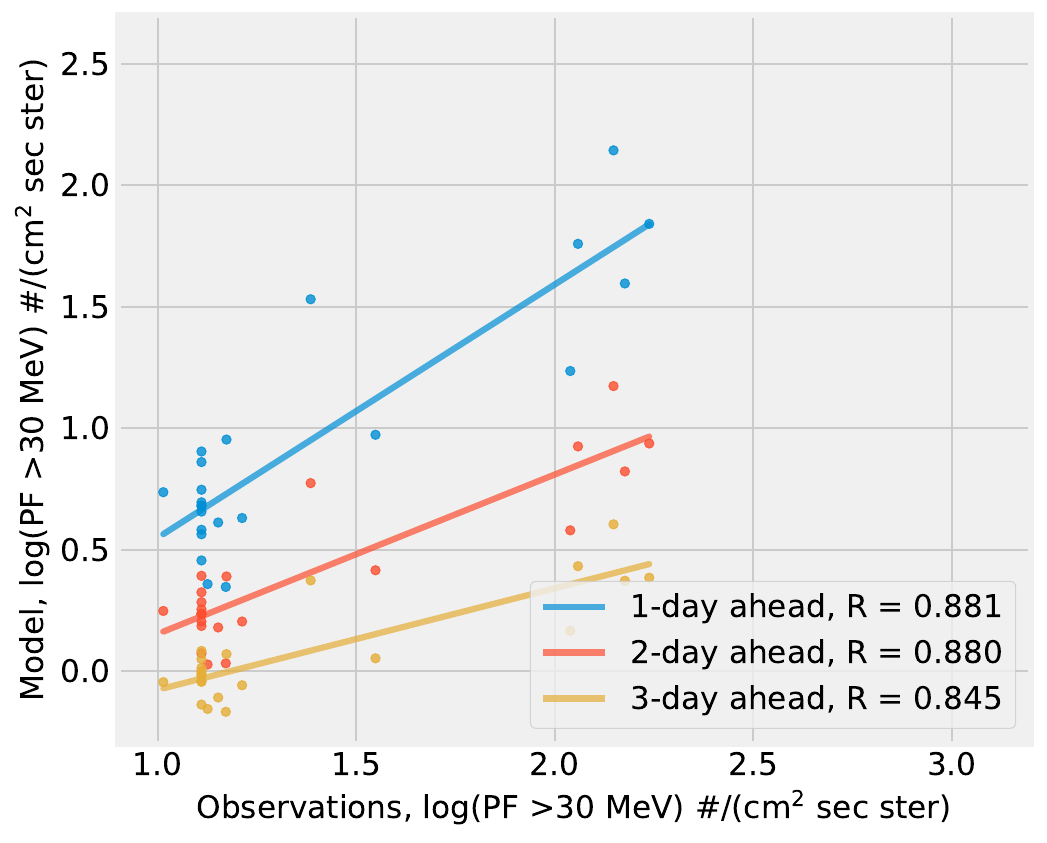}
    \end{subfigure}
    \begin{subfigure}
         \centering
         \includegraphics[width=0.4\textwidth]{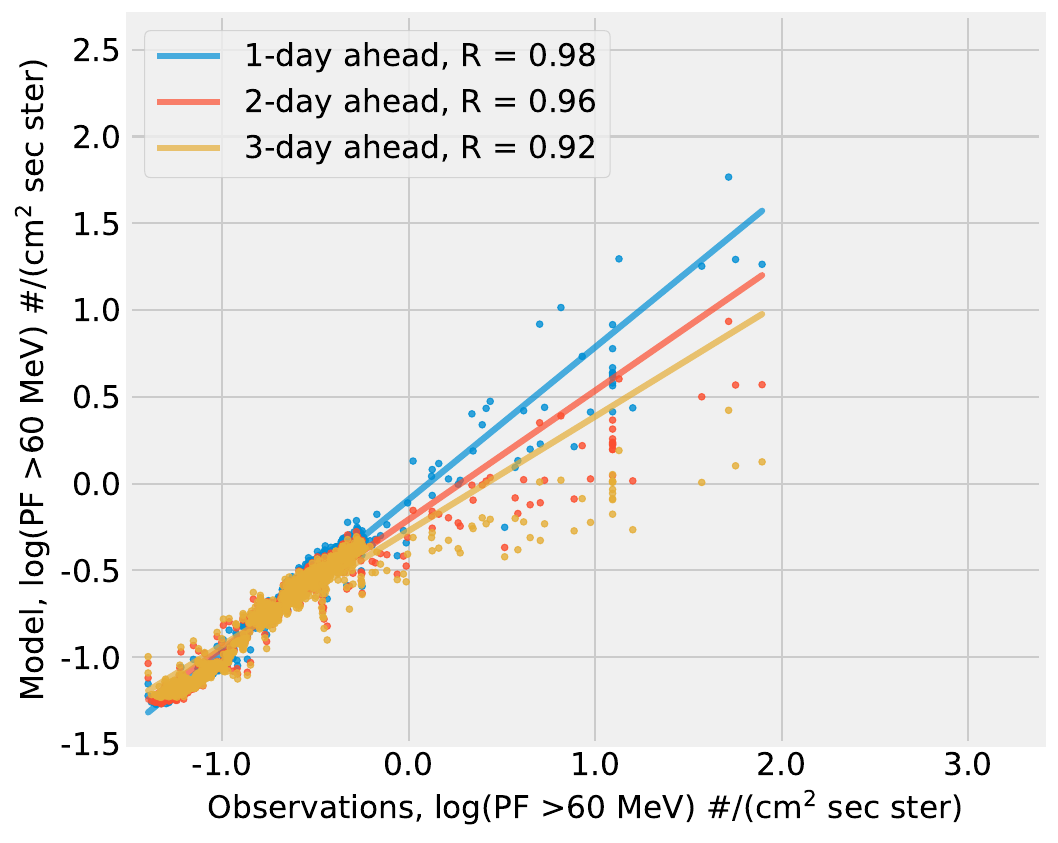}
    \end{subfigure}
    \begin{subfigure}
         \centering
         \includegraphics[width=0.4\textwidth]{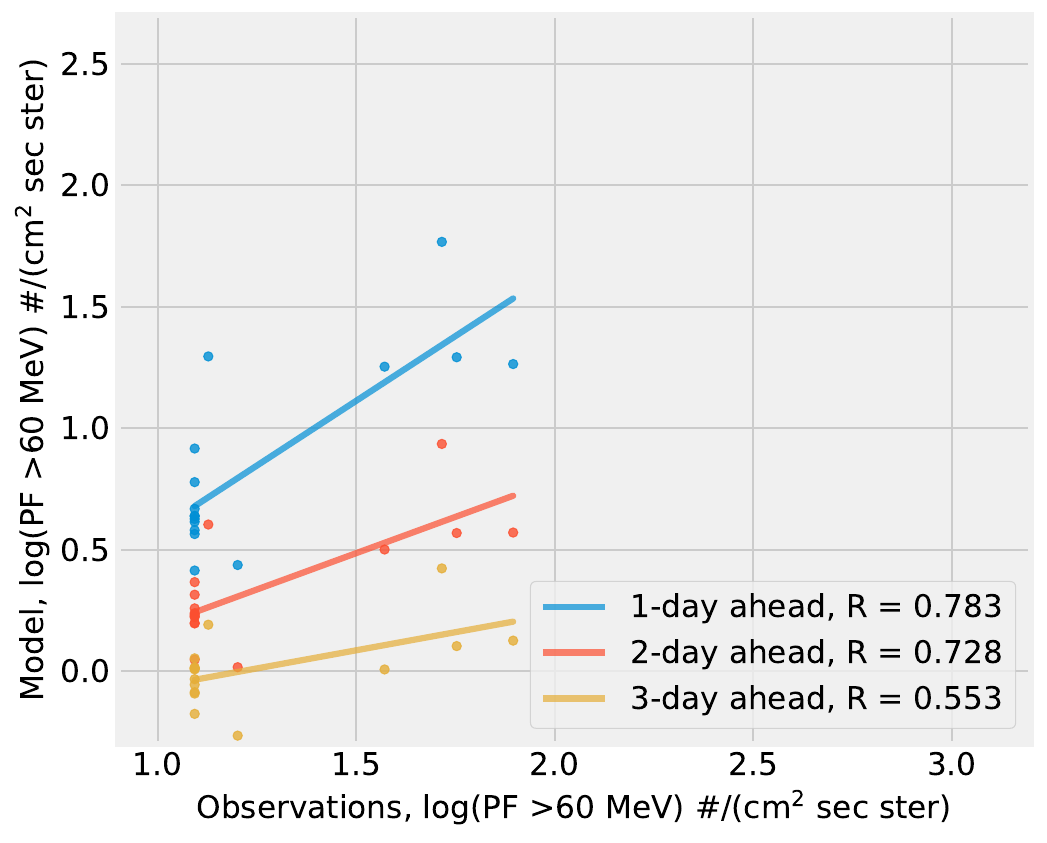}
    \end{subfigure}
\caption{Same as Figure~\ref{fig_model_vs_obs_valset} but for the test set.}
\label{fig_model_vs_obs_tstset}
\end{figure}

We found that, overall, the models performed very well. The $R$ correlation was $>$0.9 for all points of the validation set across the forecasting windows for the 3 energy channels. The $R$ correlation was $>$0.7 for the observations points $\geq$10 pfu as well. However, the correlation between the modeled data and the observations exhibited a decline as the forecast horizon increased, in accordance with the anticipated result.
To confirm the validity of the models, we performed the same correlation analysis between the modeled data and the observations of the out-of-sample test set (Fig.~\ref{fig_model_vs_obs_tstset}), which was not given to the model. Again, we found a high correlation across the forecasting windows for the 3 energy channels. 
The points were more dispersed between 1 and 1.5 on the x-axis, which reflected in a bit lower correlation. This might be a limitation in the current version of the model between that range of SEP fluxes since the models underestimated the flux values within that range across all energy channels, possibly due to the relatively smaller training samples with fluxes above 10 pfu compared with the majority of the data.

\begin{figure}[h!]
    \centerline{\includegraphics[width=\textwidth]{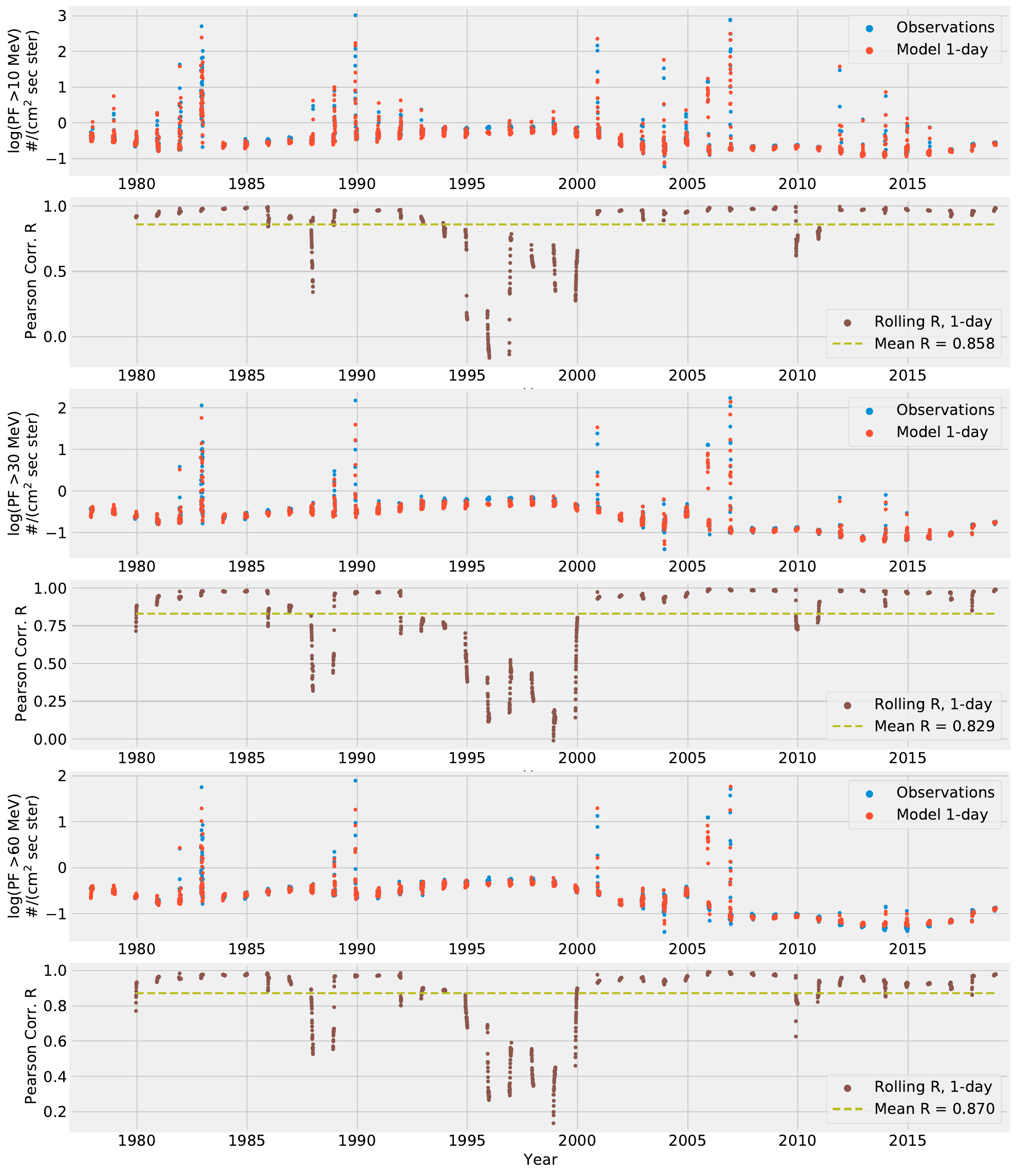}}
    \caption{Comparison between the model outputs and observations of the test set for the 3 energy channels. In addition to the rolling-mean window correlation for 1-day ahead predictions.}
\label{fig_crosscorr_tstset}
\end{figure}

In order to see the temporal variation of the correlation between the modeled data and the observations, we applied a rolling window of 90 steps (3 months × 30 days/month = 1 season) that shows the seasonal variation of the correlation, as shown in Figure~\ref{fig_crosscorr_tstset}.  
Here, we show only the 1-day ahead predictions for the test set, for the 3 energy channels. 
We observe drops in the correlation factor synchronized with the transition between solar cycles (e.g., particularly between $\sim$1995 - 2000, which represents the declining phase of the solar cycle 22 and the rising phase of the solar cycle 23). This could be related to the fact that the low SEP fluxes during quiet times are more random and thus more difficult to forecast \citep{feynman1990solar, gabriel1990periodicities, rodriguez2010east, xapsos2012periods}.

During periods of low solar activity, the forecasting of low SEP fluxes becomes more challenging due to their increased randomness. This difficulty arises from the reduced occurrence of conventional SEP drivers, such as solar flares and CMEs. Studies have suggested that the most significant solar eruptions tend to happen shortly before or after the solar cycle reaches its maximum \citep{vsvestka1995}. Additionally, sporadic increases in solar activity have been observed \citep{kane2011}, which might contribute to the diminished correlations observed in our research.
There is clearly some factor that is influencing the correlation during certain periods where there are no or only weak SEP events. However, it is not obvious which physical phenomena are the cause rather than, for instance, some artifact of the data. Understanding the interplay between these factors and their influence on SEP fluxes during periods of reduced solar activity remains a critical area of research. It would be interesting to find what is reducing the correlations, thus more investigation is needed.

Overall, the modeled data was correlated the most with observations at $>$60 MeV, then the second rank was for the $>$10 MeV channel, and the third rank was for the $>$30 MeV channel. That could be related to the relatively larger extent of drops in correlation at the $>$30 MeV channel.
The decline in correlation at the $>$30 MeV channel is consistent with the findings of \citet{le2017}.
A summary of the performance results of the models for both the validation set and test set is presented in Table~\ref{table_performance}.

\begin{figure}[htp]
    \centerline{\includegraphics[width=\textwidth]{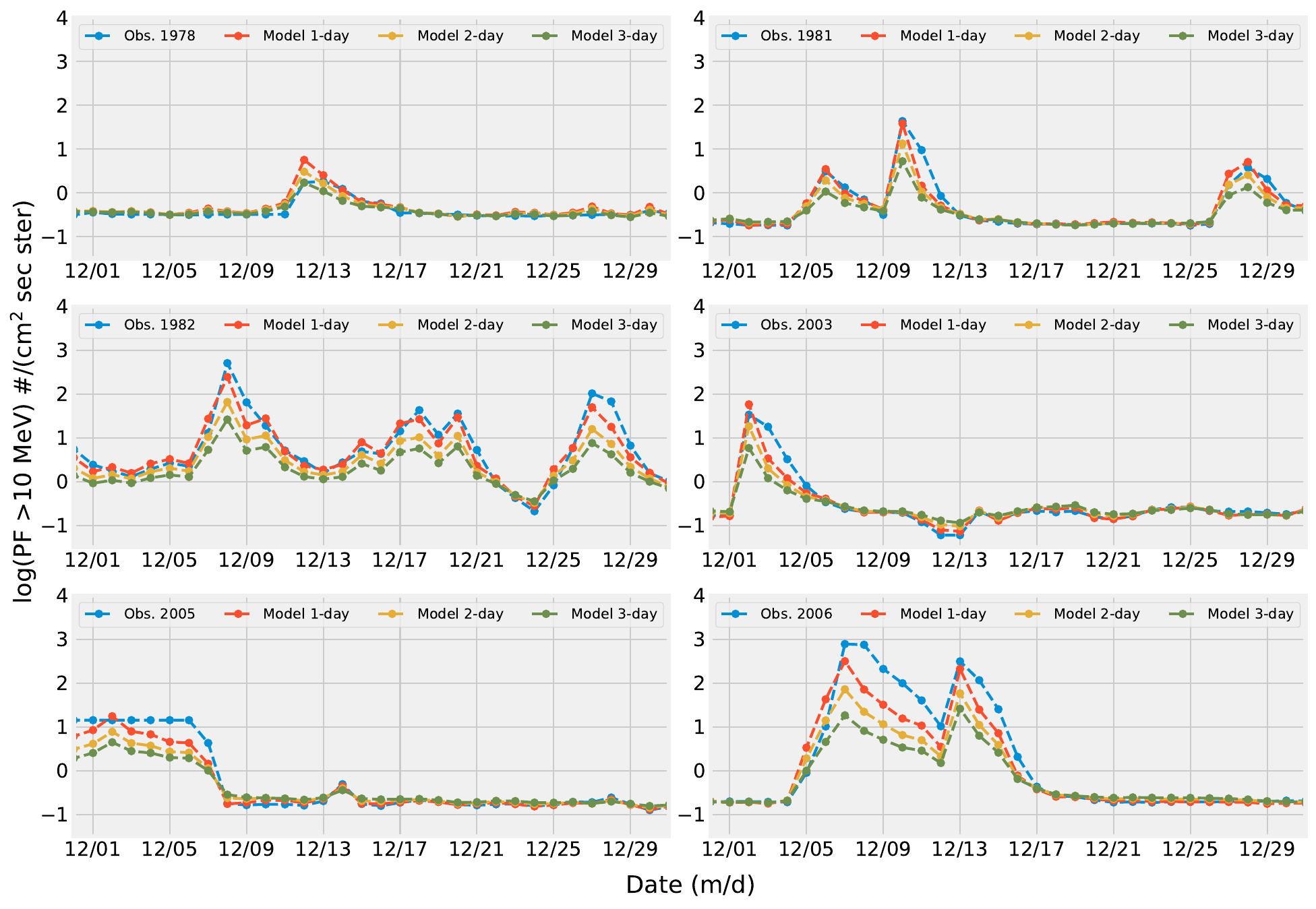}}
    \caption{The model's forecasts for the out-of-sample testing set for the $>$10 MeV channel are shown at forecast horizons of 1 day, 2 days, and 3 days ahead, using samples of data from December in selected years mentioned in the top-left side of the plots.}
\label{fig_examples_pf10_tstset}
\end{figure}

\begin{figure}[htp]
    \centerline{\includegraphics[width=\textwidth]{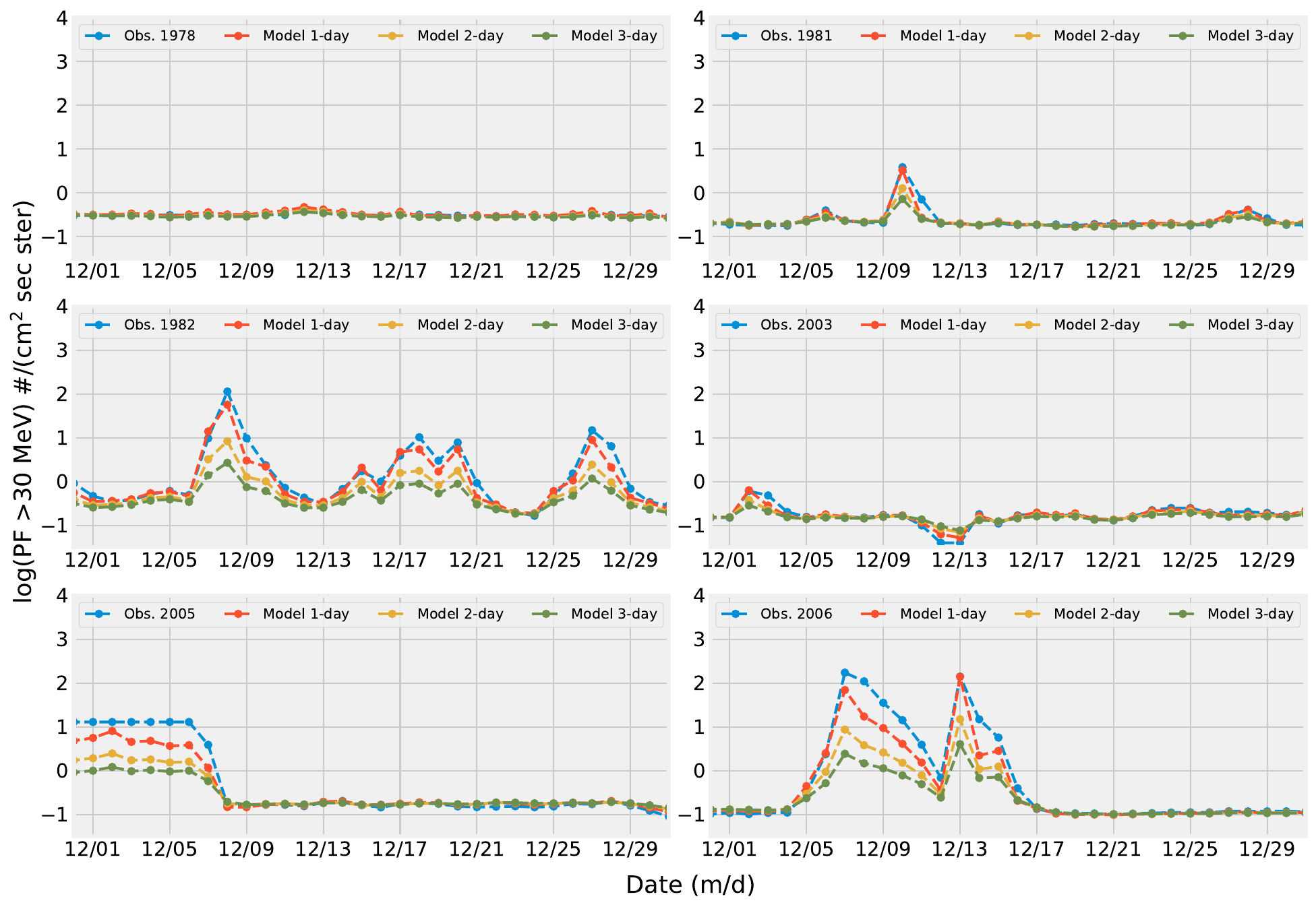}}
    \caption{The model's forecasts for the out-of-sample testing set for the $>$30 MeV channel are shown at forecast horizons of 1 day, 2 days, and 3 days ahead, using samples of data from December in selected years mentioned in the top-left side of the plots.}
\label{fig_examples_pf30_tstset}
\end{figure}

\begin{figure}[htp]
    \centerline{\includegraphics[width=\textwidth]{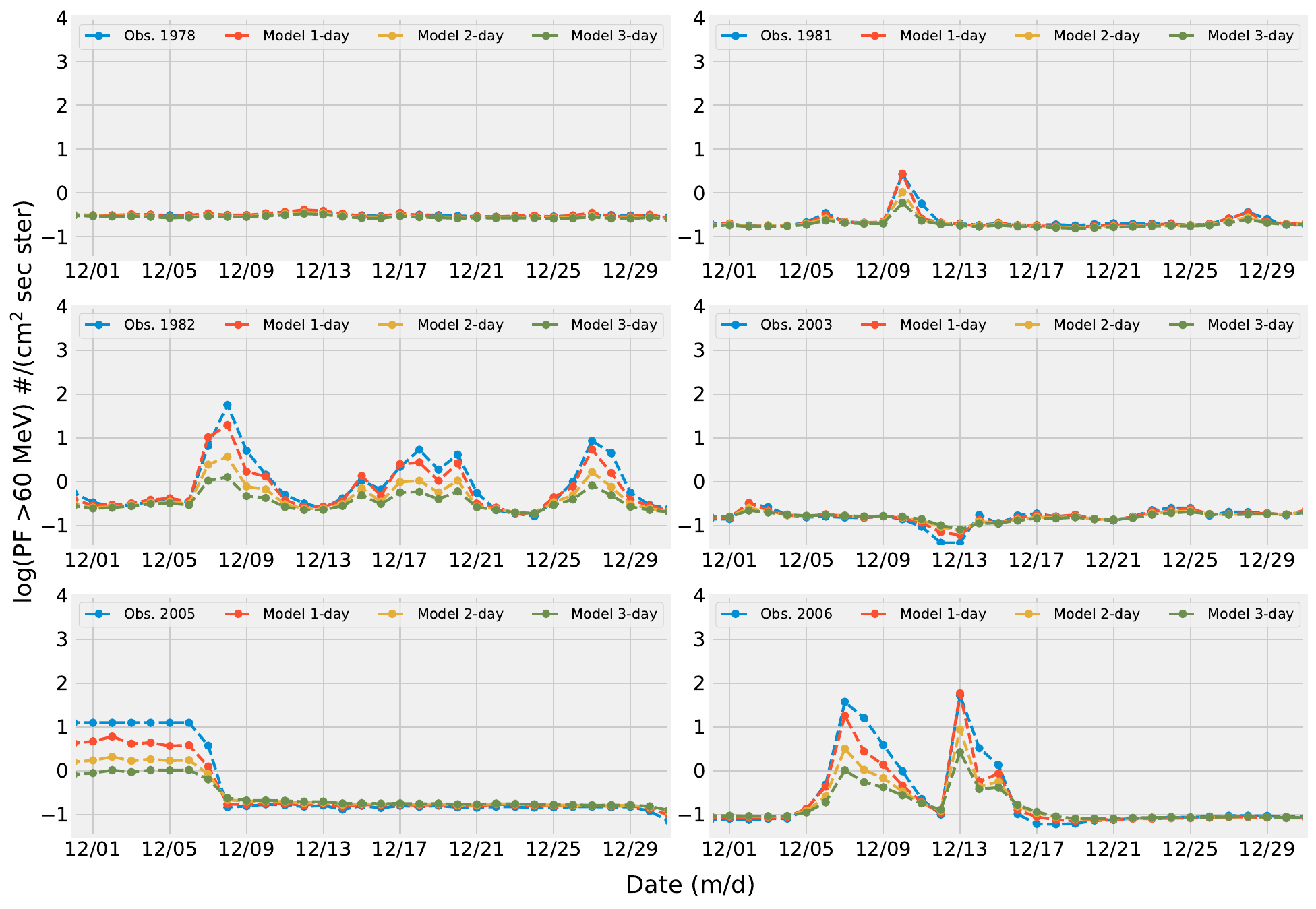}}
    \caption{The model's forecasts for the out-of-sample testing set for the $>$60 MeV channel are shown at forecast horizons of 1 day, 2 days, and 3 days ahead, using samples of data from December in selected years mentioned in the top-left side of the plots.}
\label{fig_examples_pf60_tstset}
\end{figure}

From the visual inspection of the test set examples (Fig.~\ref{fig_examples_pf10_tstset}, ~\ref{fig_examples_pf30_tstset}, and~\ref{fig_examples_pf60_tstset}), we found that the predicted onset time, the peak time, and end times of SEP events were highly correlated with those of the observations, which implies that the model captured the temporal variations, as well as the trends in SEP flux.

\begin{table}[htp]
\centering
\caption{Summary of the performance results of the models for the validation and test sets.}
\begin{tabular}{lccccccccl}
\hline
\multicolumn{10}{c}{\textbf{Validation Set}}                                                                                                                 \\ \hline
       & \multicolumn{3}{c}{log PF \textgreater{}10 MeV} & \multicolumn{3}{c}{log PF \textgreater{}30 MeV} & \multicolumn{3}{c}{log PF \textgreater{}60 MeV} \\ \hline
Model Loss   & \multicolumn{3}{c}{0.0016}                      & \multicolumn{3}{c}{0.0010}                      & \multicolumn{3}{c}{0.0009}                      \\
Model Metric & \multicolumn{3}{c}{0.0329}                      & \multicolumn{3}{c}{0.0232}                      & \multicolumn{3}{c}{0.0218}                      \\ \hline
       & 1-Day          & 2-Day          & 3-Day         & 1-Day          & 2-Day          & 3-Day         & 1-Day    & 2-Day   & \multicolumn{1}{c}{3-Day}  \\ \hline
MAE    & 0.061          & 0.091          & 0.125         & 0.053          & 0.079          & 0.098         & 0.052    & 0.069   & 0.086                      \\
MSE    & 0.013          & 0.028          & 0.054         & 0.010          & 0.031          & 0.055         & 0.009    & 0.027   & 0.047                      \\
RMSE   & 0.114          & 0.168          & 0.233         & 0.098          & 0.176          & 0.234         & 0.097    & 0.164   & 0.217                      \\
MAPE   & 22.156         & 28.104         & 34.721        & 13.039         & 18.590         & 22.735        & 10.036   & 13.994  & 16.731                     \\ \hline
\multicolumn{10}{c}{\textbf{Test Set}}                                                                                                                       \\ \hline
       & \multicolumn{3}{c}{log PF \textgreater{}10 MeV} & \multicolumn{3}{c}{log PF \textgreater{}30 MeV} & \multicolumn{3}{c}{log PF \textgreater{}60 MeV} \\ \hline
Model Loss   & \multicolumn{3}{c}{0.0014}                      & \multicolumn{3}{c}{0.0011}                      & \multicolumn{3}{c}{0.0010}                      \\
Model Metric & \multicolumn{3}{c}{0.0333}                      & \multicolumn{3}{c}{0.0283}                      & \multicolumn{3}{c}{0.0250}                      \\ \hline
       & 1-Day          & 2-Day          & 3-Day         & 1-Day          & 2-Day          & 3-Day         & 1-Day    & 2-Day   & \multicolumn{1}{c}{3-Day}  \\ \hline
MAE    & 0.072          & 0.099          & 0.125         & 0.053          & 0.088          & 0.107         & 0.045    & 0.066   & 0.081                      \\
MSE    & 0.015          & 0.030          & 0.050         & 0.009          & 0.029          & 0.048         & 0.007    & 0.020   & 0.034                      \\
RMSE   & 0.121          & 0.172          & 0.224         & 0.094          & 0.170          & 0.218         & 0.082    & 0.141   & 0.184                      \\
MAPE   & 30.135         & 37.498         & 48.139        & 20.599         & 34.300         & 40.803        & 12.358   & 20.504  & 25.305                     \\ \hline
\end{tabular}
\label{table_performance}
\end{table}

To get further insight into the model's performance, we conducted an assessment of various skill scores, including True Positive (TP), True Negative (TN), False Positive (FP), and False Negative (FN). Additionally, skill score ratios such as Probability of Detection (POD), Probability of False Detection (POFD), False Alarm Rate (FAR), Critical Success Index (CSI), True Skill Statistic (TSS), and Heidke Skill Score (HSS). Detailed descriptions of these skill scores can be found in Appendix~\ref{skillscores_appendix}.
To extract individual SEP events from the test dataset, we implemented a threshold-based clustering algorithm. This algorithm uses the NOAA/SWPC warning threshold value of 10 pfu for the E $\geq$10 MeV channel. Upon analysis, we identified the number of detected SEP events for each output forecasting window and calculated the skill scores (Table~\ref{table_skillscores}). In the true test set, we identified 12 SEP events.

The evaluation of the model revealed notable trends as the length of the output forecasting window increased. The POD and CSI exhibited a declining pattern, indicating a reduced ability of the model to accurately detect and capture positive events (SEP occurrences) as the forecasting horizon extended further into the future. This suggests that the model's performance in identifying and capturing true positive instances diminishes with longer forecasting windows. Moreover, the POFD demonstrated an increasing trend, indicating an elevated rate of false positive predictions as the forecasting horizon lengthened. The model's propensity to generate false alarms rose with the lengthening forecasting window, leading to incorrect identification of non-events as positive events. Consequently, the TSS and HSS exhibited decreasing values, signifying a deterioration in the model's overall skill in accurately capturing and distinguishing between positive and negative instances. Overall, our skill scores are comparable with those reported by previous studies (Table~\ref{table_skillscores_comparison}). Although the UMASEP model does better than ours (i.e., has a higher POD), our FAR is much lower, thus, making fewer false alarms than the UMASEP model.

\begin{table}[htp]
\centering
\caption{Confusion matrix for the energy channel $\geq$10 MeV predictions in the test set.}
\label{table_skillscores}
\begin{tabular}{lccccccccccc}
\hline
E \textgreater{}10 MeV & No. events & TP & TN   & FP & FN \\ \hline
1-day ahead            & 15         & 21 & 1441 & 2  & 13 \\ \hline
2-day ahead            & 13         & 14 & 1441 & 2  & 20 \\ \hline
3-day ahead            & 5          & 5  & 1443 & 0  & 29 \\ \hline
\end{tabular}
\end{table}

\begin{table}[htp]
\centering
\caption{Comparing the skill scores with previous models. The dashed entries mean the data is unavailable (\citet{whitman22} for more details).}
\label{table_skillscores_comparison}
\begin{tabular}{lccccccccc}
\hline
\multicolumn{1}{c}{\textbf{Model}}       &            & \textbf{POD} & \textbf{FAR} & \textbf{TSS} & \textbf{HSS} & \textbf{POFD} & \textbf{CSI} & \multicolumn{1}{l}{\textbf{Accuracy}} & \multicolumn{1}{l}{\textbf{Precision}} \\ \hline
\multirow{3}{*}{Our BiLSTM model}        & 1-Day      & 0.618        & 0.087        & 0.531        & 0.732        & 0.001         & 0.583        & 0.99                                  & 0.913                                  \\ \cline{2-10} 
                                         & 2-Day      & 0.412        & 0.125        & 0.287        & 0.553        & 0.001         & 0.389        & 0.985                                 & 0.875                                  \\ \cline{2-10} 
                                         & 3-Day      & 0.147        & 0            & 0.147        & 0.252        & 0             & 0.147        & 0.980                                 & 1                                      \\ \hline
\multicolumn{2}{l}{UMASEP-10 \citep{Nunez_2011}}           & 0.822        & 0.219        & ---          & ---          & ---           & ---          & ---                                   & ---                                    \\ \hline
\multicolumn{2}{l}{PCA \citep{Papaioannou_2018}}     & 0.587        & 0.245        & ---          & 0.65         & ---           & ---          & ---                                   & ---                                    \\ \hline
\multicolumn{2}{l}{SPARX \citep{Dalla_2017}}         & 0.5          & 0.57         & ---          & ---          & 0.32          & 0.3          & ---                                   & ---                                    \\ \hline
\multicolumn{2}{l}{SPRINTS \citep{engell_2017}}      & 0.56         & 0.34         & ---          & 0.58         & ---           & ---          & ---                                   & ---                                    \\ \hline
\multicolumn{2}{l}{REleASE \citep{malandraki_2018}} & 0.63         & 0.3          & ---          & ---          & ---           & ---          & ---                                   & ---                                    \\ \hline
\end{tabular}
\end{table}
\section{Conclusions}
\label{S_conclusion}
Forecasting the SEP flux is a crucial task in heliophysics since it affects satellite operations, astronaut safety, and ground-based communication systems. It is a challenging task due to its non-linear, non-stationary, and complex nature. Machine learning techniques, particularly neural networks, have shown promising results in predicting SEP flux.
In this study, we developed and trained BiLSTM neural network models to predict the daily-averaged integral flux of SEP at 1-day, 2-day, and 3-day ahead, for the energy channels $>$10 MeV, $>$30 MeV, and $>$60 MeV.
We used a combination of solar and interplanetary magnetic field indices from the OMNIWeb database for the past 4 solar cycles as input to the model.
We compared the models with baseline models and evaluated them using the Huber loss and the error metrics in Appendix~\ref{eval_appendix}.

The data windowing method we used, based on the MIMO strategy, eliminates the need to feed the output forecast as input back into the model and that allows to do forecasting relatively far into the future while maintaining decent results (e.g., the MSE is ranging between 0.007 and 0.015 for 1-day forecasting in the test set, compared to an MSE of 0.236 for a persistence model. See Table~\ref{table_performance}).
The results show that the model can make reasonably accurate predictions given the difficulty and complexity of the problem.
The MSE was ranged between 0.009 and 0.055 for the validation set, and between 0.007 and 0.05 for the test set.
The correlations between the observations and predictions were $>$0.9 for the validation and test sets (Fig.~\ref{fig_model_vs_obs_valset} and Fig.~\ref{fig_model_vs_obs_tstset}).
Nevertheless, the mean temporal correlation was $\sim$0.8 for the test set (Fig.~\ref{fig_crosscorr_tstset}).
Although our models performed well, we observed a relatively large discrepancy between the predictions and the observations in the $>$30 MeV energy band.

The findings of this study underscore the challenges encountered by the forecasting model in accurately predicting SEP data over longer time periods. As the length of the output forecasting window increased, the model's ability to detect true positives and its overall skill in differentiating positive and negative instances diminished. Additionally, the model displayed an elevated rate of false negative predictions, indicating an increased tendency to generate misses as the forecasting horizon extended. These results highlight the importance of carefully considering the appropriate forecasting window length for SEP data to ensure the model's optimal performance. Our skill scores generally align with those from previous works (Table~\ref{table_skillscores_comparison}). There are variations in the metrics' values across different studies, highlighting the complexities and nuances associated with each study. Nevertheless, it is important to acknowledge that the statistical significance of the results in this study is limited due to data averaging. Future studies should consider incorporating hourly data, as this is likely to result in a greater number of identified events.
The model can provide short-term predictions, which can be used to anticipate the behavior of the near-Earth space environment. These predictions have important implications for space weather forecasting, which is essential for protecting satellites, spacecraft, and astronauts from the adverse effects of solar storms.

Multiple techniques exist for identifying the optimal combination of hidden layers and neurons for a given task such as empirical methods, parametric methods, and the grid search cross-validation method, which we will explore in future work.
The observed reduction in correlation necessitates further investigation to determine its origin, whether stemming from tangible causal factors or potential aberrations within the model or data.
We plan to expand upon this work by performing short-term forecasting using hourly-averaged data. This extension will involve integrating additional relevant features such as the location and area of active regions and coronal holes on the Sun.

BiLSTM networks are particularly useful for tasks involving sequential data such as timeseries forecasting. Given their capacity to handle input sequences in both directions in time and capture long-term dependencies, they are valuable in a broad range of applications. Nonetheless, one should carefully consider their data requirements and computational complexity before adopting them.
Our results emphasize that the use of deep learning models in forecasting tasks in heliophysics are promising and encouraging, as pointed out by \citet{zhang_2022}.

This work is a stepping stone towards real-time forecasting of SEP flux based on the public-available datasets. As an extension, we are currently working on developing a set of models that deliver near-real time prediction of SEP fluxes at multiple energy bands, multiple forecasting windows, with hourly-averaged data resolution, with a more sophisticated model architecture, as well as more features that address the state of solar activity more comprehensively. 
We plan to extend the analysis to include more recent data from solar cycle 25, in order to improve the accuracy of the models.
In conclusion, our study highlights the potential of using BiLSTM neural networks for forecasting SEP integral fluxes. Our models provide a promising approach for predicting the near-Earth space environment too, which is crucial for space weather forecasting and ensuring the safety of our space assets. Our findings contribute to the growing body of literature on the applications of deep learning techniques in heliophysics and space weather forecasting. 
\begin{acknowledgements}
    We thank the referees for their feedback. We acknowledge using data from the GSFC/SPDF OMNIWeb interface. This work is supported by the Modeling and ObServational Integrated Investigations of Coronal Solar Eruptions (MOSAIICS) project, funded under contract KP-06-DV-8/18.12.2019 to the Institute of Astronomy and National Astronomical Observatory (NAO), the Bulgarian Academy of Sciences (BAS), under the National Scientific Program “VIHREN” in Bulgaria. 
\end{acknowledgements}
\bibliography{jswsc}
\appendix
\section{Terminologies}
\label{terminologies_appendix}
In this section, we introduce the main concepts related to machine learning which are presented in this paper.

\begin{itemize}[label=\textbullet]
    \item \textbf{Cross-validation}: A technique used to evaluate the performance of a machine learning model by dividing the data into subsets and assessing the model on different combinations of these subsets.

    \item \textbf{Input Horizon}: The number of previous time steps considered as input to a model for time series forecasting. It represents the length of the historical sequence used for predictions.
    
    \item \textbf{Batch Size}: The number of samples processed together in a single iteration of the training algorithm. It affects training speed and memory requirements.
    
    \item \textbf{Updating the Model's Weights}: The process of adjusting the parameters of a neural network based on training data to minimize the difference between predicted and true outputs. The model's weights represent the parameters that are learned during the training process.
    
    \item \textbf{Loss}: A function that quantifies the difference between predicted and actual outputs. It guides the optimization process during training.
    
    \item \textbf{Minimum Validation Loss}: The lowest value achieved by the loss function on a validation dataset during training. It indicates the most accurate predictions on unseen data.
    
    \item \textbf{Overfitting}: When a model performs well on training data but fails to generalize to unseen data due to memorizing training examples instead of learning underlying patterns.
    
    \item \textbf{Learning Rate}: A hyperparameter that determines the step size at each iteration of the optimization algorithm during training. It affects learning speed and convergence. A high learning rate can cause the training process to converge quickly, but it may also result in overshooting the optimal solution or getting stuck in a suboptimal solution. On the other hand, a very low learning rate can make the training process slow, and may struggle to find the optimal solution.
    
    \item \textbf{Reducing the learning rate when the validation loss stops improving}: This concept involves adjusting the learning rate dynamically during the training process. When the validation loss reaches a plateau or stops improving, it indicates a suboptimal point. By reducing the learning rate, the model can take smaller steps in weight space, potentially finding a better solution. This technique, known as learning rate scheduling or learning rate decay, is commonly used to fine-tune the model's performance.
    
    \item \textbf{Patience}: A parameter used in training to determine the number of epochs to wait for an improvement in validation loss before stopping the training process.
    
    \item \textbf{Patience Parameter of 7}: In the context of early stopping, training will be stopped if the validation loss does not improve for 7 consecutive epochs.
    
    \item \textbf{Adam Optimizer}: A popular optimization algorithm in deep learning that combines Adaptive Gradient Algorithm (AdaGrad) and Root Mean Square Propagation (RMSprop) to achieve efficient optimization.
    
    \item \textbf{Optimal Architecture}: The best configuration of a neural network, including the number of layers, neurons, and other choices, for optimal performance on a specific task.
    
    \item \textbf{Hyperparameters}: Parameters set before training a model that control the learning algorithm's behavior, such as learning rate, batch size, and activation functions.
    
    \item \textbf{Layer}: A building block of a neural network that performs specific operations on input data. Includes input, hidden, output, fully connected, convolutional, recurrent, activation, and dropout layers. Here is a description for each layer:
    
    \begin{itemize}
        \item \textbf{Input Layer}: The first layer of a neural network that receives raw input data. It passes the input to subsequent layers for further processing. The number of nodes in the input layer is determined by the dimensionality of the input data.
    
        \item \textbf{Hidden Layers}: Intermediate layers between the input and output layers. They perform computations on the input data and capture higher-level representations or abstractions. Hidden layers are not directly exposed to the input or output.
        
        \item \textbf{Output Layer}: The final layer of a neural network that produces model predictions or outputs based on computations from preceding layers. The number of neurons in the output layer depends on the problem being solved, such as regression or classification.
        
        \item \textbf{Fully Connected Layer (Dense Layer)}: Each neuron in this layer is connected to every neuron in the previous layer. It allows information flow between all neurons, enabling complex relationships to be learned.
        
        \item \textbf{Convolutional Layer}: Commonly used in Convolutional Neural Networks (CNNs) for analyzing grid-like data, such as images. It applies convolution operations using filters or kernels to learn spatial patterns or features.
        
        \item \textbf{Recurrent Layer}: Used in Recurrent Neural Networks (RNNs) to process sequential data. These layers have feedback connections that allow information to be passed from one step to the next, capturing temporal dependencies and maintaining memory of past inputs.
        
        \item \textbf{Activation Layer}: Applies a non-linear function to the output of a layer, introducing non-linearity into the neural network. Activation functions like Sigmoid, Hyperbolic Tangent (tanh), or Rectified Linear Unit (ReLU) determine neuron outputs based on weighted inputs.
        
        \item \textbf{Dropout Layer}: A regularization technique commonly used in deep learning models. It randomly sets a fraction of outputs from the previous layer to zero during training, preventing overfitting and improving generalization.
    \end{itemize}
    
    Layers play a crucial role in the information processing and learning capabilities of neural networks. The arrangement and combination of different layers determine the network's architecture and ultimately its ability to solve specific tasks.
    
    \item \textbf{Stateful}: A property of Recurrent Neural Networks (RNNs) where the hidden state is preserved between consecutive inputs, allowing the network to have memory.
    
    \item \textbf{Neuron}: A computational unit in a neural network that receives input, applies weights, and passes the result through an activation function to produce an output.
    
    \item \textbf{Hidden Neuron}: A neuron in a hidden layer of a neural network that performs intermediate computations.
    
    \item \textbf{Callback Function}: A function used during model training to perform specific actions at certain points or conditions, such as saving the best model, adjusting learning rates, or early stopping.
    
    \item \textbf{\textit{LearningRateScheduler} Callback Function}: A function used in training to dynamically adjust the learning rate at specific points based on a predefined schedule or function. It improves training efficiency and convergence by allowing the model to make finer adjustments as it approaches the optimal solution.
\end{itemize}
\section{Mathematical Representation of the LSTM NN Model}
\label{bilstm_appendix}
The computations inside one LSTM cell can be described by the following formulas \citep{ihianle_2020}:
\begin{subequations}
    \begin{gather}
        f_t = \sigma(W_f x_t + U_f h_{t-1} + b_f)\\ 
        i_t = \sigma(W_i x_t + U_i h_{t-1} + b_i)\\ 
        \tilde{C_t} = \tanh(W_c x_t + U_c h_{t-1} + b_c)\\
        C_t = f_t \odot C_{t-1} + i_t \odot \tilde{C_t}\\
        o_t = \sigma(W_o x_t + U_o h_{t-1} + b_o)\\
        h_t = o_t \odot \tanh(C_t)
    \end{gather}
    \label{eq_lstm}
\end{subequations}
where $x_t$ is input data at time $t$. The input gate $i_t$ determines which values from the updated cell states (candidate values) $\tilde{C_t}$ should be added to the cell state. It also takes into account the current input $x_t$ and the previous output $h_{t-1}$, and is passed through a sigmoid activation function. 
$\tilde{C_t}$ represent the candidate values that are added to the cell state at time $t$. 
The forget gate activation vector $f_t$ at time step $t$, which determines how much of the previous cell state should be retained. 
The cell state $C_t$ at time $t$ is updated based on the forget gate, input gate, and candidate values. 
The output gate $o_t$ at time $t$ determines how much of the cell state should be output. 
The output vector $h_t$ at time $t$ is calculated based on the cell state and the output gate values.
$h_{t-1}$ is the output vector at the previous time step $t-1$. 
$W_f, W_i, W_c, W_o$ are the weight matrices for the input vector $x_t$. 
$U_f, U_i, U_c, U_o$ are the weight matrices for the output vector $h_{t-1}$.
$b_f, b_i, b_c, b_o$ are the bias vectors. 
The symbol $\odot$ denotes a pointwise multiplication. 
The sigmoid function $\sigma$ is used as the activation function for the gate vectors, and the hyperbolic tangent function $\tanh$ is used for the candidate values and the output vector.
\section{Evaluation Metrics}
\label{eval_appendix}
To evaluate the model performance, we used the following equations:
\begin{subequations}
\begin{gather}
    L_\delta (y, \hat{y}) = 
    \begin{cases}
    \frac{1}{2} (y - \hat{y})^2, & \text{if } |y - \hat{y}| \leq \delta,\\
    \delta (|y - \hat{y}| - \frac{1}{2} \delta), & \text{otherwise}
    \label{eq_huber}
    \end{cases}
    \\
    MSE = \frac{1}{N} \sum_{i=1}^{N} (y_i - \hat{y_i})^2\\ 
    MAE = \frac{1}{N} \sum_{i=1}^{N} |y_i - \hat{y_i}|\\ 
    RMSE = \sqrt{\frac{\sum_{i=1}^{N}(y_i - \hat{y_i})^2}{N}}\\ 
    MAPE = \frac{1}{N} \sum_{i=1}^{N} |\frac{y_i - \hat{y_i}}{\hat{y_i}}|\\ 
    R = \frac{\sum_{i=1}^n (y_i - \bar{y}) (y_i - \bar{y})}{\sqrt{\sum_{i=1}^n (y_i - \bar{y})^2} \sqrt{\sum_{i=1}^n (y_i - \bar{y})^2}}
\end{gather}
\label{eq_metrics}
\end{subequations}
where $y$ is the true value, $\hat{y}$ is the predicted value, and $\delta$ is a threshold in the Huber loss function that controls the trade-off between the mean squared error (MSE) and the mean absolute error (MAE). In this paper, it was set to 0.1, which was selected based on several experiments.

MSE is the mean squared error, which measures the difference between predicted and actual values by calculating the average of squared differences. It provides a measure of the average squared magnitude of the errors in your forecasts, which can be useful in penalizing larger errors more heavily than smaller errors. 

MAPE is the mean absolute percentage error, which measures the difference between predicted and actual values by calculating the average of absolute differences. It provides a measure of the average magnitude of the errors, allowing to evaluate the overall accuracy of your forecasts. 

RMSE is the root mean squared error, which measures the difference between predicted and actual values by taking the square root of the average of squared differences. It provides a measure of the accuracy of the forecasts in the same units as the original data, allowing to evaluate the magnitude of errors in the same scale as the data. 

MAPE is the mean absolute percentage error, which measures the accuracy of a forecast by calculating the average of absolute percentage errors. It provides a measure of the accuracy of the forecasts in percentage terms, allowing to evaluate the magnitude of errors relative to the actual values.
MSE, MAE, RMSE, and MAPE are often used in regression analysis to assess the accuracy of the model's predictions.

Finally, $R$ is the Pearson correlation coefficient, which measures the strength and direction of the relationship between two continuous variables, and can provide an indication of the extent to which changes in one variable may be related to changes in the other.
\section{Model Configuration}
\label{config_appendix}
The configurations for the ML models shown in Figure~\ref{fig_benchmark} and their performance on the validation set and the test set for the SEP integral flux $\geq$10 MeV are presented in Table~\ref{table_models_config}.
The batch size was set to be 64 and the number of training epochs was set to be 100. 
The \textit{EarlyStopping} callback function, with a \textit{patience} of 10, is used to help prevent overfitting during the training process by stopping training when the monitored metric has stopped improving for a certain number of epochs.
The \textit{patience} parameter controls how many epochs the training will continue without improvement before it is stopped. This is useful because if the validation loss stops getting better, the model has probably overfitted the training data and is not generalizing effectively to new data. By stopping the training early, we can avoid wasting time and resources on further training that is unlikely to improve the model's performance.

We used the \textit{ModelCheckpoint} callback function to save the best weights of the model during training so that they can be reused later.
The \textit{LearningRateScheduler} callback function allows to dynamically adjust the learning rate of the model during training using a function passed to it that will be called at the beginning of each epoch, and it should return the desired learning rate for that epoch. It can be useful when training deep neural networks, as it allows for a higher learning rate in the early stages of training when the model is still far from convergence, and a lower learning rate as the model approaches convergence, which can help it to converge more accurately. The downside might be the longer training time.

\begin{table}[h!]
\caption{Configuration of the ML model. (1) refers to the error value for 1-day forecasting. Same for (2) refers to 2-day forecasting, and (3) for 3-day forecasting. $^*$In the 1D-CNN layer, 32 filters, a kernel size of 5, and strides of 1 were used.}
\label{table_models_config}
\resizebox{\textwidth}{!}{%
\begin{tabular}{cccccccccccccccc}
\hline
\multirow{2}{*}{\begin{tabular}[c]{@{}c@{}}Model\\ Architecture\end{tabular}} & \multirow{2}{*}{\begin{tabular}[c]{@{}c@{}}No. of\\ Hidden Layers\end{tabular}} & \multirow{2}{*}{\begin{tabular}[c]{@{}c@{}}No. of\\ Hidden Neurons\end{tabular}} & \multirow{2}{*}{\begin{tabular}[c]{@{}c@{}}Activation\\ Function\end{tabular}} & \multirow{2}{*}{\begin{tabular}[c]{@{}c@{}}Batch\\ Size\end{tabular}} & \multirow{2}{*}{\begin{tabular}[c]{@{}c@{}}Learning\\ Rate\end{tabular}} & \multirow{2}{*}{Epochs} & \multirow{2}{*}{\begin{tabular}[c]{@{}c@{}}Callbacks\\ Functions\end{tabular}} & \multicolumn{4}{c}{Validation Set}                                                                                                                                                                                                                                                                                  & \multicolumn{4}{c}{Testing Set}                                                                                                                                                                                                                                                                                     \\ \cline{9-16} 
                                                                              &                                                                                 &                                                                                  &                                                                                &                                                                       &                                                                          &                         &                                                                                & MAE                                                                       & MSE                                                                       & RMSE                                                                      & MAPE                                                                            & MAE                                                                       & MSE                                                                       & RMSE                                                                      & MAPE                                                                            \\ \hline
Linear                                                                        & --                                                                             & --                                                                              & --                                                                            & 64                                                                    & 0.001                                                                    & 100                     & EarlyStopping                                                                  & 0.312                                                                     & 0.141                                                                     & 0.376                                                                     & 87.883                                                                          & 0.143                                                                     & 0.045                                                                     & 0.211                                                                     & 60.689                                                                          \\ \hline
\begin{tabular}[c]{@{}c@{}}Dense\\ ML\end{tabular}                            & 2                                                                               & 32                                                                               & ReLU                                                                           & 64                                                                    & 0.001                                                                    & 100                     & EarlyStopping                                                                  & \begin{tabular}[c]{@{}c@{}}0.262 (1)\\ 0.275 (2)\\ 0.290 (3)\end{tabular} & \begin{tabular}[c]{@{}c@{}}0.118 (1)\\ 0.138 (2)\\ 0.166 (3)\end{tabular} & \begin{tabular}[c]{@{}c@{}}0.344 (1)\\ 0.372 (2)\\ 0.407 (3)\end{tabular} & \begin{tabular}[c]{@{}c@{}}132.580 (1)\\ 132.004 (2)\\ 129.288 (3)\end{tabular} & \begin{tabular}[c]{@{}c@{}}0.400 (1)\\ 0.395 (2)\\ 0.392 (3)\end{tabular} & \begin{tabular}[c]{@{}c@{}}0.281 (1)\\ 0.286 (2)\\ 0.294 (3)\end{tabular} & \begin{tabular}[c]{@{}c@{}}0.530 (1)\\ 0.535 (2)\\ 0.542 (3)\end{tabular} & \begin{tabular}[c]{@{}c@{}}238.898 (1)\\ 234.704 (2)\\ 230.896 (3)\end{tabular} \\ \hline
\begin{tabular}[c]{@{}c@{}}Simple\\ RNN\end{tabular}                          & 2                                                                               & 32                                                                               & Tanh                                                                           & 64                                                                    & 0.001                                                                    & 100                     & \begin{tabular}[c]{@{}c@{}}EarlyStopping\\ ModelCheckpoint\end{tabular}        & \begin{tabular}[c]{@{}c@{}}0.143 (1)\\ 0.171 (2)\\ 0.264 (3)\end{tabular} & \begin{tabular}[c]{@{}c@{}}0.035 (1)\\ 0.063 (2)\\ 0.118 (3)\end{tabular} & \begin{tabular}[c]{@{}c@{}}0.187 (1)\\ 0.251 (2)\\ 0.343 (3)\end{tabular} & \begin{tabular}[c]{@{}c@{}}70.990 (1)\\ 68.694 (2)\\ 72.505 (3)\end{tabular}    & \begin{tabular}[c]{@{}c@{}}0.178 (1)\\ 0.171 (2)\\ 0.200 (3)\end{tabular} & \begin{tabular}[c]{@{}c@{}}0.052 (1)\\ 0.071 (2)\\ 0.084 (3)\end{tabular} & \begin{tabular}[c]{@{}c@{}}0.228 (1)\\ 0.266 (2)\\ 0.289 (3)\end{tabular} & \begin{tabular}[c]{@{}c@{}}69.624 (1)\\ 78.075 (2)\\ 67.416 (3)\end{tabular}    \\ \hline
\begin{tabular}[c]{@{}c@{}}Stateful\\ RNN\end{tabular}                        & 3                                                                               & 32                                                                               & Tanh                                                                           & 64                                                                    & 1.58e$^{-4}$                                                             & 100                     & \begin{tabular}[c]{@{}c@{}}LearningRateScheduler\\ EarlyStopping\end{tabular}  & \begin{tabular}[c]{@{}c@{}}0.203 (1)\\ 0.305 (2)\\ 0.349 (3)\end{tabular} & \begin{tabular}[c]{@{}c@{}}0.060 (1)\\ 0.131 (2)\\ 0.173 (3)\end{tabular} & \begin{tabular}[c]{@{}c@{}}0.244 (1)\\ 0.362 (2)\\ 0.416 (3)\end{tabular} & \begin{tabular}[c]{@{}c@{}}56.390 (1)\\ 81.028 (2)\\ 82.819 (3)\end{tabular}    & \begin{tabular}[c]{@{}c@{}}0.155 (1)\\ 0.223 (2)\\ 0.223 (3)\end{tabular} & \begin{tabular}[c]{@{}c@{}}0.039 (1)\\ 0.079 (2)\\ 0.084 (3)\end{tabular} & \begin{tabular}[c]{@{}c@{}}0.197 (1)\\ 0.281 (2)\\ 0.289 (3)\end{tabular} & \begin{tabular}[c]{@{}c@{}}59.988 (1)\\ 71.679 (2)\\ 64.159 (3)\end{tabular}    \\ \hline
\begin{tabular}[c]{@{}c@{}}Stateful\\ LSTM\end{tabular}                       & 3                                                                               & 32                                                                               & Tanh                                                                           & 64                                                                    & 1.58e$^{-4}$                                                             & 100                     & EarlyStopping                                                                  & \begin{tabular}[c]{@{}c@{}}0.095 (1)\\ 0.151 (2)\\ 0.174 (3)\end{tabular} & \begin{tabular}[c]{@{}c@{}}0.021 (1)\\ 0.048 (2)\\ 0.076 (3)\end{tabular} & \begin{tabular}[c]{@{}c@{}}0.146 (1)\\ 0.220 (2)\\ 0.275 (3)\end{tabular} & \begin{tabular}[c]{@{}c@{}}40.335 (1)\\ 48.937 (2)\\ 55.662 (3)\end{tabular}    & \begin{tabular}[c]{@{}c@{}}0.098 (1)\\ 0.134 (2)\\ 0.166 (3)\end{tabular} & \begin{tabular}[c]{@{}c@{}}0.020 (1)\\ 0.042 (2)\\ 0.071 (3)\end{tabular} & \begin{tabular}[c]{@{}c@{}}0.141 (1)\\ 0.205 (2)\\ 0.267 (3)\end{tabular} & \begin{tabular}[c]{@{}c@{}}41.781 (1)\\ 57.860 (2)\\ 68.025 (3)\end{tabular}    \\ \hline
\begin{tabular}[c]{@{}c@{}}Stateful\\ Bi-LSTM\end{tabular}                    & 3                                                                               & 32                                                                               & Tanh                                                                           & 64                                                                    & 1.58e$^{-4}$                                                             & 100                     & EarlyStopping                                                                  & \begin{tabular}[c]{@{}c@{}}0.149 (1)\\ 0.190 (2)\\ 0.249 (3)\end{tabular} & \begin{tabular}[c]{@{}c@{}}0.043 (1)\\ 0.074 (2)\\ 0.120 (3)\end{tabular} & \begin{tabular}[c]{@{}c@{}}0.207 (1)\\ 0.272 (2)\\ 0.347 (3)\end{tabular} & \begin{tabular}[c]{@{}c@{}}58.151 (1)\\ 60.154 (2)\\ 67.988 (3)\end{tabular}    & \begin{tabular}[c]{@{}c@{}}0.170 (1)\\ 0.211 (2)\\ 0.229 (3)\end{tabular} & \begin{tabular}[c]{@{}c@{}}0.049 (1)\\ 0.090 (2)\\ 0.108 (3)\end{tabular} & \begin{tabular}[c]{@{}c@{}}0.221 (1)\\ 0.300 (2)\\ 0.329 (3)\end{tabular} & \begin{tabular}[c]{@{}c@{}}71.059 (1)\\ 92.727 (2)\\ 87.049 (3)\end{tabular}    \\ \hline
\begin{tabular}[c]{@{}c@{}}1D-CNN\\ LSTM\end{tabular}                         & 3                                                                               & 32 (5,1)$^*$                                                                     & \begin{tabular}[c]{@{}c@{}}ReLU\\ Tanh\end{tabular}                            & 64                                                                    & 1.58e$^{-4}$                                                             & 100                     & EarlyStopping                                                                  & \begin{tabular}[c]{@{}c@{}}0.108 (1)\\ 0.146 (2)\\ 0.177 (3)\end{tabular} & \begin{tabular}[c]{@{}c@{}}0.027 (1)\\ 0.051 (2)\\ 0.078 (3)\end{tabular} & \begin{tabular}[c]{@{}c@{}}0.165 (1)\\ 0.226 (2)\\ 0.279 (3)\end{tabular} & \begin{tabular}[c]{@{}c@{}}41.164 (1)\\ 47.512 (2)\\ 53.087 (3)\end{tabular}    & \begin{tabular}[c]{@{}c@{}}0.098 (1)\\ 0.138 (2)\\ 0.156 (3)\end{tabular} & \begin{tabular}[c]{@{}c@{}}0.023 (1)\\ 0.047 (2)\\ 0.067 (3)\end{tabular} & \begin{tabular}[c]{@{}c@{}}0.151 (1)\\ 0.217 (2)\\ 0.259 (3)\end{tabular} & \begin{tabular}[c]{@{}c@{}}51.732 (1)\\ 68.376 (2)\\ 69.338 (3)\end{tabular}    \\ \hline
\end{tabular}%
}
\end{table}

All the calculations and model runs were implemented under the framework of TensorFlow 2.3.0 \citep{singh_2020} in Python 3.6.13. 
The models were executed on Ubuntu 20.04.1 LTS OS with 4 × GPUs (NVIDIA GeForce RTX 2080 Ti, 11019 MiB, 300 MHz).
According to the Keras API guide \citep{keras2017}, the requirements to use the cuDNN implementation are the activation function must be set to \textit{tanh} and the recurrent activation must be set to \textit{sigmoid}.
We also set the seed number to 7 across all the model runs to maintain reproducibility.

Stateful RNNs can be difficult to work with when using callbacks in Keras because their hidden state must be manually managed across mini-batch updates. When training a stateful RNN in Keras, the hidden state is carried over from the previous epoch and can cause problems with certain callbacks, such as \textit{EarlyStopping} or \textit{ModelCheckpoint}. To work around this issue, one can use stateless RNNs or manually reset the hidden state at the end of each epoch, but this can be complex and prone to errors.
\section{Skill Scores}
\label{skillscores_appendix}
Skill scores and ratios are commonly used in evaluating the performance of classification models, particularly in binary classification tasks. They provide insights into the model's ability to correctly predict positive and negative instances. Here is a brief description of each skill score and ratio, along with their formulas:

\begin{itemize}
    \item \textbf{True Positive (TP)}: The number of data points or intervals correctly identified as positive by the model. It represents instances where both the model and the ground truth indicate the presence of an event.
    
    \item \textbf{True Negative (TN)}: The number of intervals correctly identified as negative by the model. It represents instances where both the model and the ground truth indicate the absence of an event.

    \item \textbf{False Positive (FP)}: The number of intervals incorrectly identified as positive by the model. It occurs when the model predicts an event, but the ground truth indicates its absence.

    \item \textbf{False Negative (FN)}: The number of intervals incorrectly identified as negative by the model. It occurs when the model fails to detect an event that the ground truth indicates its presence.

    \item \textbf{Accuracy}: Represents the proportion of correct predictions out of total predictions.
    \begin{equation}
        Accuracy = \frac{TP + TN}{TP + TN + FP + FN}
    \end{equation}

    \item \textbf{Precision}: Represents the proportion of positive predictions that are actually positive.
    \begin{equation}
        Precision = \frac{TP}{TP + FP}
    \end{equation}
    
    \item \textbf{Probability of Detection (POD) or Recall}: Represents the model's ability to correctly identify positive instances.
    \begin{equation}
        POD = \frac{TP}{TP + FN}
    \end{equation}

    \item \textbf{Probability of False Detection (POFD)}: Measures the model's tendency to falsely predict positive instances when the ground truth indicates their absence.
    \begin{equation}
        POFD = \frac{FP}{FP + TN}
    \end{equation}

    \item \textbf{False Alarm Rate (FAR)}: Indicates the ratio of false positive predictions to the total number of positive instances.
    \begin{equation}
        FAR = \frac{FP}{FP + TP}
    \end{equation}

    \item \textbf{Critical Success Index (CSI)}: Measures the model's ability to correctly predict both positive and negative instances.
    \begin{equation}
        CSI = \frac{TP}{TP + FP + FN}
    \end{equation}

    \item \textbf{True Skill Statistic (TSS)}: Takes into account both the model's ability to detect positive instances and its ability to avoid false alarms.
    \begin{equation}
        TSS = POD - FAR
    \end{equation}

    \item \textbf{Heidke Skill Score (HSS)}: Evaluates the model's performance by comparing it with random chance. It takes into account the agreement between the model's predictions and the observed data, considering both true positive and true negative predictions.
    \begin{equation}
        HSS = \frac{TP + TN - C}{T - C}
    \end{equation}
    where
    \begin{subequations}
        \begin{gather*}
            T = TP + TN + FP + FN\\
            C = \frac{(TP+FP)(TP+FN) + (TN+FP)(TN+FN)}{T}
        \end{gather*}
    \end{subequations}
\end{itemize}
\end{document}